\documentclass[12pt]{article}
\usepackage[utf8]{inputenc}
\usepackage{graphicx, subfigure, float}
\usepackage[margin=1.0in]{geometry}
\usepackage{amsmath, amsfonts, amssymb, amsthm} % AMS packages
\usepackage{float} 
\usepackage{booktabs}
\usepackage{siunitx}
\usepackage{appendix}
\usepackage{hyperref} % for clickable url
\graphicspath{{./Figures/}}

\title{Strain-Rate- and Line-Length-Dependent Screw Dislocation Glide Mechanisms in BCC Refractory Metals and Alloys}
\author{Subhendu Chakraborty $^1$, Liang Qi $^1$}
\date{%
    $^1$ Department of Materials Science and Engineering, University of Michigan, Ann Arbor, Michigan 48109, USA \\%
    \today
}

\begin{document}

\maketitle
\begin{abstract}

Plastic flow in body-centered cubic (BCC) metals and dilute/concentrated alloys is governed by the motion of $\langle111\rangle$ screw dislocations, whose glide is often impeded by cross-kinks (jogs). While existing strengthening models typically treat depinning as defect-assisted cutting or dislocation bowing, the combined strain-rate and dislocation-line-length dependence of cross-kink stability and effective obstacle spacing remains insufficiently resolved at the atomistic scale. Here, we combine conventional molecular dynamics and strain-boost hyperdynamics to investigate screw-dislocation glide in pure Nb and Mo, dilute Nb--Mo alloys, and equiatomic NbMo at 300~K over strain rates from $10^{3}$ to $10^{7}\ \mathrm{s^{-1}}$ and dislocation line lengths from 15 to 50~nm. We first demonstrate that low-strain-rate simulations require sufficiently long dislocation lines to capture consistent cross-kink behavior and strength-determining pinning events. Using the 50~nm configurations, we show that cross-kinks form not only in concentrated alloys but also in pure BCC metals, with their stability governed by the relative rates of kink nucleation and migration on primary and cross-slip planes, which differ between Nb- and Mo-rich systems due to distinct core structures and non-Schmid responses. At high strain rates, depinning proceeds predominantly via vacancy--interstitial cluster formation. In contrast, at low strain rates and long line lengths, alternative pathways emerge, including lateral cross-kink migration, three-dimensional forward--backward cross-slip, and prismatic loop formation. The effective obstacle spacing controlling the critical resolved shear stress therefore emerges from coupled thermodynamic roughening and kinetic evolution. These findings highlight the intrinsically rate-, length-, and chemistry-dependent nature of screw-dislocation strengthening in BCC alloys.

\end{abstract}

\section{Introduction}\label{sec:Intro}

\begin{figure}[!htb]
\centering
    \includegraphics[width=0.95\textwidth]{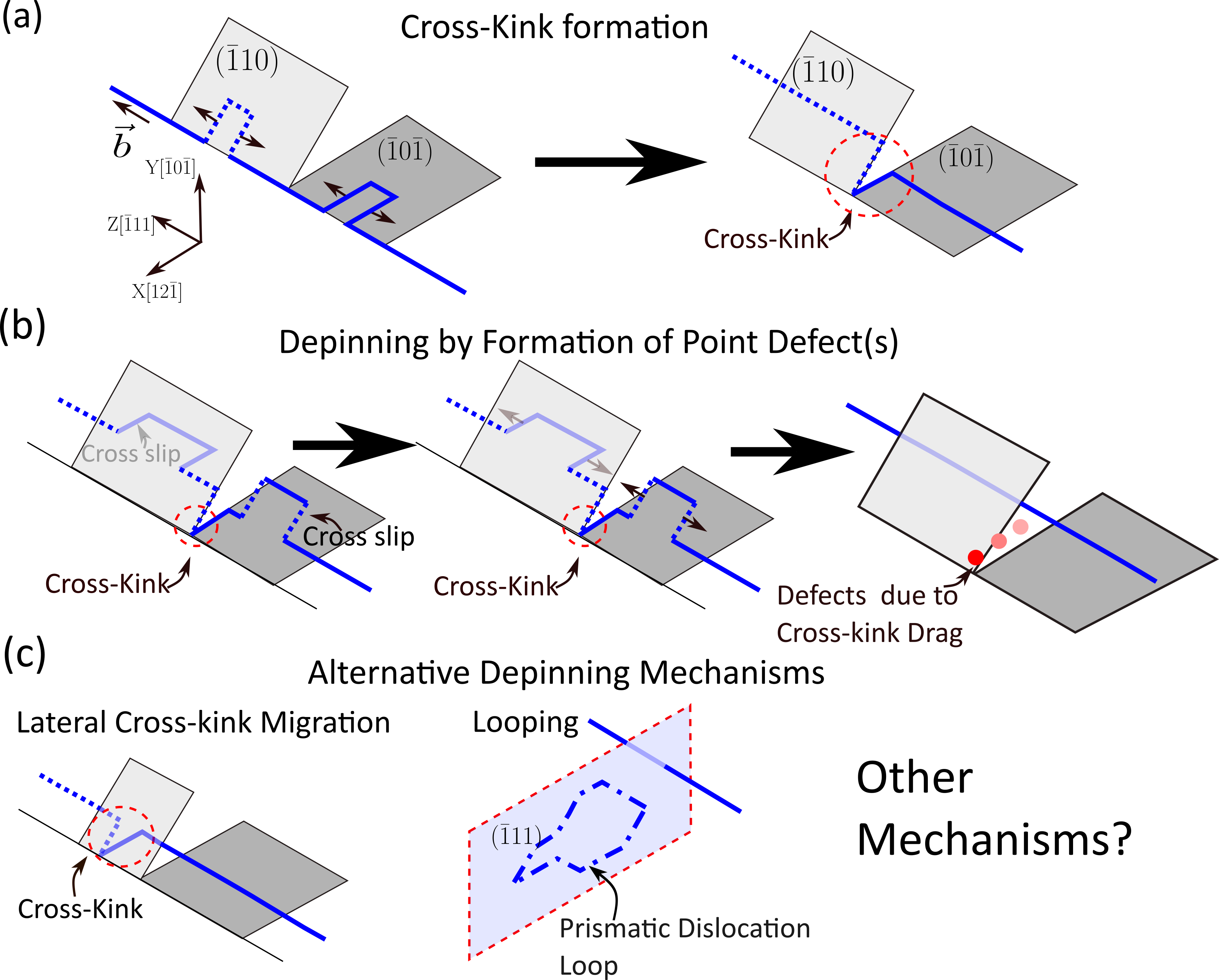}
    \caption{Schematic illustration of cross-kink formation and depinning mechanisms for $\langle111\rangle$ screw dislocations in BCC alloys. (a) Formation of a cross-kink (jog) arising from the intersection of kinks nucleated on different slip planes along the same screw-dislocation line. (b) In classical strengthening models, depinning of a cross-kink is commonly assumed to occur via defect-assisted cutting, accompanied by the formation of vacancy or interstitial debris~\cite{suzuki1980solid,Rao2019ModelingSolutionHardening,Maresca2020TheoryScrewDislocation}. (c) The present work reveals additional depinning pathways, including lateral cross-kink migration, loop emission, and other three-dimensional dislocation-line rearrangements, that become operative at low strain rates and for long dislocation line lengths. Solid and dashed blue lines denote dislocation segments on the $(\bar{1}0\bar{1})$ and $(\bar{1}10)$ slip planes, respectively; this convention is adopted consistently in all subsequent dislocation-line segment representations.}
\label{fig:GraphicalAbstract}
\end{figure}

Plastic deformation in body-centered cubic (BCC) metals and alloys is primarily governed by the motion of $\langle111\rangle$ screw dislocations, whose mobility is controlled by thermally activated double-kink nucleation followed by kink migration across a periodic Peierls potential~\cite{argon2007strengthening}. In single-element BCC metals, such as niobium (Nb), molybdenum (Mo), and tungsten (W), as well as their dilute alloys, this mechanism largely determines the yield strength and low-temperature plastic flow~\cite{trinkle2005chemistry,groger2008multiscaleI,groger2008multiscaleII,groger2008multiscaleIII,hu2017solute,Ghafarollahi2020TheoryDoublekinkNucleation,Ghafarollahi2021TheoryKinkMigration}. However, when the atomic environment becomes compositionally heterogeneous, as in refractory complex concentrated alloys (RCCAs) and refractory high-entropy alloys (RHEAs), where all principal elements are present at non-dilute concentrations, screw dislocations move through a chemically complex energy landscape that lacks a single, well-defined Peierls barrier; consequently, the classical picture of simple kink nucleation and migration on unique slip planes may no longer apply~\cite{wang2020multiplicity,wang2022hierarchical,Zhou2023ModelsDislocationGlide}. Multiple double kinks may nucleate on nearby segments of the same dislocation line; in particular, owing to the absence of close-packed crystal planes and the nonplanar structure of screw dislocation cores in BCC metals~\cite{duesbery1998plastic,vitek2004core,Cai2004DislocationCoreEffects,dezerald2016plastic}, these kink formation events can occur on distinct $\{110\}$ or $\{112\}$ slip systems. The subsequent intersection of kinks on different planes leads to the formation of cross-kinks (also called jogs), as illustrated in Fig.~\ref{fig:GraphicalAbstract}(a), and act as strong pinning points for subsequent kink migration~\cite{Suzuki1979SolidSolutionHardening,suzuki1980solid,Hattendorf92}, thereby contributing significantly to dislocation strengthening, as demonstrated by theoretical models, atomistic simulations, and mesoscale approaches~\cite{suzuki1980solid,Rao2019ModelingSolutionHardening,Maresca2020TheoryScrewDislocation,Rao2021TheorySolidSolution,Ghafarollahi2022ScrewcontrolledStrengthBCC,yin2021atomistic,cai2001kinetic,Zhou2021CrosskinksControlScrew}.

A quantitative description of screw-dislocation-controlled strengthening in non-dilute BCC alloys therefore hinges on two closely related issues: (i) how cross-kinks (or jogs) are overcome during glide, and (ii) how the characteristic spacing between these pinning points is determined, which in turn controls the stress required for dislocation bowing or jog dragging. In Suzuki’s classical solid-solution hardening theory, jogs arise from interactions between kinks gliding on different slip planes, and plastic flow is governed by a competition between kink mobility and jog-controlled Orowan bowing~\cite{Suzuki1979SolidSolutionHardening,suzuki1980solid,Hattendorf92}. In this framework, the mean jog spacing \(2L\) is not prescribed a priori but is selected dynamically through minimization of the total flow stress, reflecting the stress-dependent nature of the dominant obstacles. Importantly, once formed, these jogs are assumed to be overcome by cutting processes associated with the formation of vacancy or interstitial point defects. Building on Suzuki’s picture, Rao \emph{et al.} extended this jog-based mechanism to RCCAs by incorporating more detailed screw-dislocation–solute core interactions, thereby refining the stress-based description of jog formation, dragging, and vacancy–interstitial dipole generation~\cite{Rao2019ModelingSolutionHardening,Rao2021TheorySolidSolution}. In parallel, Maresca and Curtin developed a statistical-mechanical framework in which solute–core interactions lead to equilibrium roughening of screw dislocations and a characteristic cross-kink spacing that scales with the intrinsic kink length \(\zeta_c\) and decreases with increasing solute concentration~\cite{Maresca2020TheoryScrewDislocation}. Subsequent extensions incorporated strain-rate and temperature effects while retaining cross-kink-controlled kinetics as the central strengthening mechanism~\cite{Ghafarollahi2022ScrewcontrolledStrengthBCC}. Despite differences in how solute–dislocation core interactions are modeled and how the characteristic spacing between pinning points is determined, these approaches are physically consistent in identifying cross-kinks or jogs as dominant obstacles and in commonly treating their depinning through defect-assisted mechanisms during the cutting processes, schematically illustrated in Fig.~\ref{fig:GraphicalAbstract}(b).

While the theoretical frameworks discussed above provide physically consistent descriptions of cross-kink–controlled strengthening, directly connecting their assumptions to atomistic behavior requires resolving the detailed mechanisms of cross-kink (or jog) formation and depinning under well-controlled thermomechanical conditions. Beyond uncertainties associated with interatomic potential fidelity and chemical short-range order (SRO) effects~\cite{yin2021atomistic,wang2023tailoring}, a fundamental limitation of atomistic simulations lies in their restricted accessible time and length scales. Conventional molecular dynamics (MD) simulations typically operate at very high strain rates ($10^{6}$--$10^{8}\ \mathrm{s^{-1}}$) and employ short dislocation line lengths, conditions under which only a limited subset of depinning pathways, most notably cross-kink failure via point-defect generation, can be readily sampled. Consequently, alternative depinning mechanisms, such as lateral cross-kink migration without point-defect formation or loop-emission processes (illustrated in Fig.~\ref{fig:GraphicalAbstract}(c)), may be underrepresented or entirely inaccessible. The relative importance of these mechanisms is expected to depend sensitively on strain rate and dislocation line length, which together control elastic line tension, kink interactions, and the time available for thermally activated rearrangements. Consistent with this view, Ji~(2020) demonstrated that key kinetic quantities, including kink diffusion coefficients and friction, exhibit a strong dependence on the dislocation line length in the simulation supercell~\cite{ji2020quantifying}. Although the finite-temperature coarse-grained framework introduced by Ji~(2022) enables simulations of micrometer-long screw dislocations and captures pronounced line-length effects~\cite{ji2022finite}, the applied strain rates remain comparable to those used in conventional MD. While kinetic Monte Carlo (kMC) approaches can access lower strain rates and longer dislocation lines~\cite{cai2001kinetic,mohri2017mechanical,Zhou2021CrosskinksControlScrew}, their reliance on predefined event catalogs and energy barriers, often parameterized from MD or simplified line-tension models, may limit their ability to capture strain-rate-dependent pathway changes in chemically rough energy landscapes.

Taken together, these considerations highlight a clear gap: a systematic atomistic investigation that simultaneously resolves strain-rate effects, dislocation line-length effects, and chemically induced cross-kink energetics within a unified simulation framework is still lacking. Motivated by this gap, the objective of the present study is to employ both conventional and accelerated MD simulations~\cite{Voter1997HyperdynamicsAcceleratedMolecular} to elucidate how solute concentration, strain rate, and dislocation line length jointly govern the formation, evolution, and depinning of cross-kink defects in pure BCC refractory metals, dilute refractory alloys, and refractory concentrated alloys. Specifically, we perform atomistic simulations of screw-dislocation motion in pure Mo and Nb, dilute Mo--Nb and Nb--Mo alloys, and equiatomic NbMo as a representative refractory concentrated alloy at room temperature over a wide range of applied strain rates ($10^{3}$--$10^{7}\ \mathrm{s^{-1}}$). Strain-boost hyperdynamics~\cite{Hara2010AdaptiveStrainboostHyperdynamics} is employed to directly access long-time dynamics at low strain rates, while systematic variation of the dislocation line length enables explicit assessment of elastic line-tension and configurational roughening effects. Computational details of both conventional MD and strain-boost hyperdynamics simulations are provided in Sec.~\ref{sec:CompMethod}.

In Sec.~\ref{sec:Results}, we demonstrate that both strain rate and dislocation line length exert a strong influence on screw-dislocation glide trajectories, cross-kink formation, and depinning mechanisms across pure, dilute, and concentrated BCC systems. As benchmarks, Sections~\ref{sec:Results_15nm} and~\ref{sec:Results_20nm} establish pronounced strain-rate-dependent cross-kink behavior at moderate dislocation lengths (15 and 20~nm), including under relatively low strain rates ($\sim10^{4}\ \mathrm{s^{-1}}$). To approach more realistic atomistic conditions, simulations with an extended dislocation line length of 50~nm are performed for pure and dilute Nb–Mo systems (Sec.~\ref{sec:Results_50nm_a}), where we illustrate chemistry-dependent differences in cross-slip propensity, non-Schmid behavior, and potential lateral cross-kink migration mechanism between Nb- and Mo-rich alloys. We then investigate equiatomic NbMo as a representative concentrated alloy with an extended dislocation line length of 50~nm over a wide range of applied strain rates ($10^{3}$--$10^{7}\ \mathrm{s^{-1}}$)(Sec.~\ref{sec:Results_50nm_b}), where the chemically rough energy landscape gives rise to stronger pinning points and more complex three-dimensional depinning pathways, including extended multi-plane cross-kink structures, forward–backward cross-slip processes, and prismatic loop formation~\cite{swinburne2016fast}. Finally, in Sec.~\ref{sec:Discussion}, we discuss the implications of these findings for physically based strengthening models in BCC metals and alloys.

\section{Computational Methods}\label{sec:CompMethod}
Classical molecular dynamics (MD) simulations provide a robust framework for investigating deformation mechanisms with full atomic resolution, but their applicability to thermally activated processes is limited by the short time scales accessible in direct integration. As a consequence, simulations of continuously deforming systems are typically conducted at high imposed strain rates, often on the order of $\dot{\varepsilon} \sim 10^{7}\ \mathrm{s^{-1}}$. To extend the accessible time scale while retaining atomic fidelity, we employ strain-boost hyperdynamics~\cite{Hara2010AdaptiveStrainboostHyperdynamics}, an accelerated MD technique that enables the efficient sampling of rare, thermally activated events under applied deformation. This approach allows us to directly probe deformation mechanisms at substantially lower effective strain rates within the MD framework.

%There are several methods that has been proposed to accelerate the time evolution of an atomistic system. implemented within Large-scale Atomic/Molecular Massively Parallel Simulator (LAMMPS) platform as the base code \cite{ThompsonLAMMPSFlexibleSimulation2022}.

\subsection{General Principle of Hyperdynamics (HD)}\label{sec:HD}

\begin{figure}[!htb]
\includegraphics[width=0.9\textwidth]{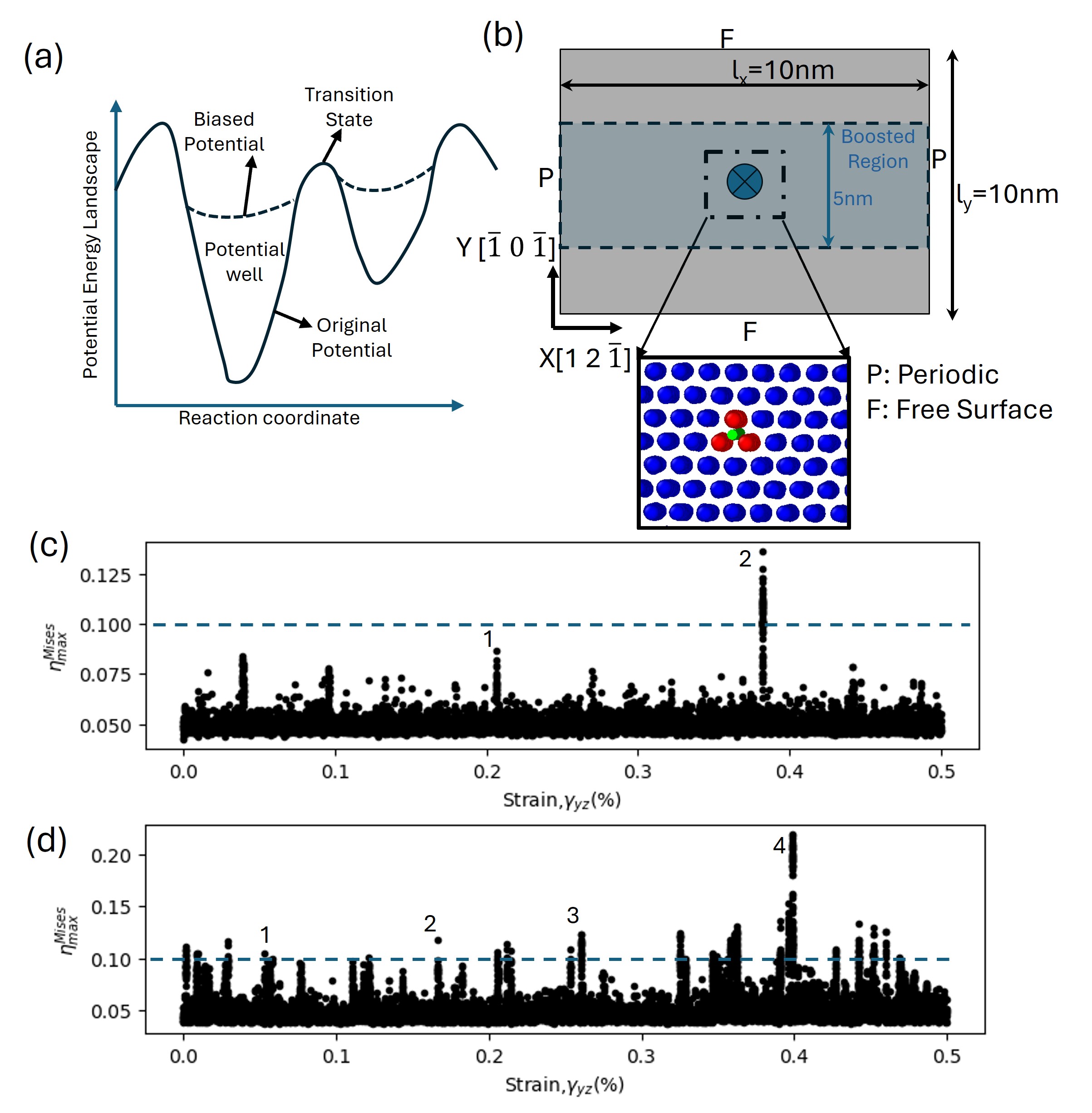}
\caption {Illustration of the hyperdynamics framework and the determination of key parameters used in this study. (a) Schematic representation of the hyperdynamics method~\cite{Voter1997HyperdynamicsAcceleratedMolecular}. The solid curve denotes the original potential-energy surface, while the dashed curve represents the biased potential with elevated basin energies and unchanged transition states. (b) Simulation supercell containing an embedded $\langle111\rangle$ screw dislocation used in the present study. The same $X$--$Y$--$Z$ coordinate system is adopted for all subsequent dislocation-structure visualizations. (c) Evolution of $\eta^{\mathrm{Mises}}_{\mathrm{max}}$ defined in Eq.~\ref{eq:StrainBoost_StoppingFunction} for a Nb single crystal with a pre-existing screw dislocation at 300~K. (d) Corresponding evolution of $\eta^{\mathrm{Mises}}_{\mathrm{max}}$ for a NbMo alloy containing a pre-existing screw dislocation at 300 K.
}
\label{fig:MethodHyperdynamics}
\end{figure}

An efficient approach for accelerating transitions between adjacent potential-energy basins in atomistic systems was introduced through the hyperdynamics (HD) method by Voter~\cite{Voter1997HyperdynamicsAcceleratedMolecular}. The central idea of hyperdynamics is to accelerate rare-event dynamics by lifting the bottoms of potential-energy wells through the introduction of a bias (or boost) potential, while leaving the transition-state regions unchanged. A schematic illustration of this concept is shown in Fig.~\ref{fig:MethodHyperdynamics}(a). In this framework, a boost potential is added to the original system potential (solid line), effectively raising the energy of the basin. The resulting biased potential (dashed line) increases the frequency of transitions between metastable states, thereby accelerating the system’s dynamics. 

The time evolution of the biased system can be rigorously related to that of the original, unbiased system using Transition State Theory (TST)~\cite{Eyring1935ActivatedComplexChemical, Truhlar1980VariationalTransitionstateTheory, Vanden-Eijnden2005TransitionStateTheory}. By combining TST with equilibrium statistical-mechanics ensemble averaging, the physical time advancement of the system under the biased potential can be expressed as
\begin{equation}
\label{eq:HD_boost}
\Delta t_{\mathrm{phy}} = \Delta t_{\mathrm{MD}} \exp\!\left(\frac{\Delta V_{\mathrm{b}}}{k_{\mathrm{B}} T}\right),
\end{equation}
where $\Delta t_{\mathrm{phy}}$ is the effective physical time corresponding to an MD time increment $\Delta t_{\mathrm{MD}}$, $\Delta V_{\mathrm{b}}$ is the applied boost potential, $T$ is the simulation temperature, and $k_{\mathrm{B}}$ is the Boltzmann constant. A comprehensive theoretical formulation and practical implementation of the hyperdynamics method can be found in Refs.~\cite{Voter1997HyperdynamicsAcceleratedMolecular, Kim2014PracticalPerspectiveImplementation}.

A central challenge in hyperdynamics is the construction of an appropriate boost potential that provides sufficient acceleration of the system dynamics while remaining consistent with the assumptions of Transition State Theory (TST). A general form of the boost potential can be written as
\begin{equation}
\label{eq:Hyperdynamics_PE}
\Delta V(\mathbf{r}) = \frac{F}{N_{\mathrm{b}}} \sum_{i} \delta V_{i},
\end{equation}
where $\delta V_{i}$ is the boost contribution associated with the $i$th atom, and $N_{\mathrm{b}}$ is the total number of atoms subject to the boost. The factor $F$ is a stopping function designed to smoothly drive the boost potential to zero as the system approaches the dividing surface (the \emph{transition state} illustrated in Fig.~\ref{fig:MethodHyperdynamics}(a)) between metastable states, thereby ensuring that the fundamental assumptions of hyperdynamics are not violated. 

Various strategies have been proposed to construct effective boost potentials that maximize time acceleration while minimizing computational overhead, including bond-based and strain-based approaches~\cite{Miron2003AcceleratedMolecularDynamics, Hara2010AdaptiveStrainboostHyperdynamics, Huang2015HyperdynamicsBoostFactor}. The method introduced by Miron and Fichthorn~\cite{Miron2003AcceleratedMolecularDynamics} is commonly referred to as \emph{bond-boost hyperdynamics}, as it employs interatomic bond lengths as collective variables to define the boost potential. In contrast, the approach developed by Hara and Li~\cite{Hara2010AdaptiveStrainboostHyperdynamics} is known as \emph{strain-boost hyperdynamics}, since the local atomic strain is used as the collective variable. In the present study, we adopt the strain-boost hyperdynamics method proposed in Ref.~\cite{Hara2010AdaptiveStrainboostHyperdynamics}. A brief description of this method is provided in the following section.

\subsection{Strain Boost Hyperdynamics} \label{sec:strain_boost}
The boost potential used in the present study is constructed as a function of the second invariant of the least-squares atomic strain, as defined in Ref.~\cite{Zimmerman2009DeformationGradientsContinuum}. With this formulation, Eq.~\eqref{eq:Hyperdynamics_PE} takes the form
\begin{subequations}
\begin{align}
\Delta V(\mathbf{r}) &= \frac{F(\eta^{\mathrm{Mises}}_{\mathrm{max}})}{N_{\mathrm{b}}}
\sum_i \delta V_i(\eta^{\mathrm{Mises}}_i), 
\label{eq:StrainBoost_PE} \\
\delta V_i(\eta^{\mathrm{Mises}}_i) &= 
V_{\mathrm{max}}\left[1 - \left(\frac{\eta^{\mathrm{Mises}}_i}{q_{\mathrm{c}}}\right)^2 \right],
\label{eq:StrainBoost_PEi}
\end{align}
\label{eq:strboost}
\end{subequations}
where $\eta^{\mathrm{Mises}}_i$ is the second invariant of the local atomic strain $\boldsymbol{\eta}_i$ of the $i$th atom, and
$\eta^{\mathrm{Mises}}_{\mathrm{max}} = \max \{\eta^{\mathrm{Mises}}_i \mid i=1,2,\ldots,N_{\mathrm{b}}\}$, with $N_{\mathrm{b}}$ denoting the number of atoms subject to the boost. The quantity $\delta V_i$ is the atomic boost potential, which depends explicitly on $\eta^{\mathrm{Mises}}_i(\mathbf{r})$. The function $F(\eta^{\mathrm{Mises}}_{\mathrm{max}})$ is a stopping function that enforces the fundamental assumptions of hyperdynamics by smoothly suppressing the boost potential as the system approaches a transition state. In the present study, the stopping function is defined as
\begin{equation}
\label{eq:StrainBoost_StoppingFunction}
F(\eta^{\mathrm{Mises}}_{\mathrm{max}})=
\begin{cases}
1 - \left(\dfrac{\eta^{\mathrm{Mises}}_{\mathrm{max}}}{q_{\mathrm{c}}}\right)^2,
& \forall~ \eta^{\mathrm{Mises}}_{\mathrm{max}} < q_{\mathrm{c}}, \\
0,
& \forall~ \eta^{\mathrm{Mises}}_{\mathrm{max}} \geq q_{\mathrm{c}} .
\end{cases}
\end{equation}

The collective variable used to construct both the boost potential and the stopping function is the local atomic strain. The procedure for computing local Lagrangian atomic strains $\boldsymbol{\eta}_i$ from atomic configurations is described in Refs.~\cite{Zimmerman2009DeformationGradientsContinuum,Hara2010AdaptiveStrainboostHyperdynamics}. To ensure frame invariance of the boost potential, the second invariant of the deviatoric part of $\boldsymbol{\eta}_i$ is employed, defined as
\begin{equation}
\label{eq:StrainBoost_etaMises}
\eta^{\mathrm{Mises}}_i \equiv 
\sqrt{\frac{1}{2}\,\mathrm{Tr}\!\left(\boldsymbol{\eta}_i - \eta^{\mathrm{hydro}}_i \mathbf{I}\right)^2},
\end{equation}
where $\mathrm{Tr}$ denotes the trace operator, $\eta^{\mathrm{hydro}}_i$ is the hydrostatic component of $\boldsymbol{\eta}_i$, and $\mathbf{I}$ is the identity tensor. The strain-based boost potential requires two parameters, $V_{\mathrm{max}}$ and $q_{\mathrm{c}}$, as defined in Eqs.~\eqref{eq:strboost} and~\eqref{eq:StrainBoost_StoppingFunction}. The procedure used to determine these parameters for a given material is discussed below.

A key prerequisite of hyperdynamics is that the boost potential $\delta V_i$ must vanish on all dividing surfaces as the system approaches a saddle point between adjacent potential-energy basins, as illustrated schematically in Fig.~\ref{fig:MethodHyperdynamics}(a). In principle, saddle points of the potential-energy surface can be identified by evaluating the gradient vector $g_i = \partial V / \partial x_i$ and the Hessian matrix $H_{ij} = \partial^2 V / (\partial x_i \partial x_j)$, where $\mathbf{x}$ is a $3N$-dimensional vector containing the atomic coordinates of a system with $N$ atoms~\cite{Voter1997HyperdynamicsAcceleratedMolecular}. Solving $g_i = 0$ yields all extrema of the energy surface, and a saddle point corresponds to an extremum at which the Hessian matrix has exactly one negative eigenvalue. However, this approach is impractical for large atomistic systems, as the potential-energy surface is not known \emph{a priori}, and locating all stationary points and evaluating the associated Hessians is computationally prohibitive.

To overcome this limitation, hyperdynamics exploits the observation that, when a system crosses a saddle point separating two potential-energy basins, it undergoes a significant local configurational change involving at least one atom and its nearest neighbors. Consequently, any state variable sensitive to such configurational changes can serve as an indicator of an impending transition. A critical value of this indicator can then be used as a threshold beyond which the boost potential is forced to zero. In the present study, the stopping function $F(\eta^{\mathrm{Mises}}_{\mathrm{max}})$ in Eq.~\eqref{eq:StrainBoost_StoppingFunction} fulfills this role, with $q_{\mathrm{c}}$ defining the critical value of $\eta^{\mathrm{Mises}}_{\mathrm{max}}$.

Substantial changes in $\eta^{\mathrm{Mises}}$ at the onset of dislocation kink nucleation or migration make it a particularly effective indicator of transition events. An alternative indicator, originally proposed for bond-boost hyperdynamics~\cite{Miron2003AcceleratedMolecularDynamics}, is the critical bond length: when a tagged bond exceeds a predefined threshold, the system is considered to be approaching a transition. In the present work, the objective is to investigate screw-dislocation migration kinetics involving kink nucleation and migration on multiple slip planes. For this problem, the strain-boost hyperdynamics approach is preferred over the bond-boost method, as kink nucleation and migration are collective processes involving clusters of atoms. The local atomic strain therefore provides a more appropriate collective variable than the maximum bond length between neighboring atoms.

\subsection{Setup of MD simulations} 
To investigate strain-rate- and line-length-dependent screw-dislocation glide mechanisms in refractory concentrated alloys, we employ the strain-boost hyperdynamics (SBHD) method implemented in the open-source MD package LAMMPS~\cite{Chakraborty2016AcceleratedMolecularDynamics,LAMMPS}. The method is applied to study the migration of $\langle111\rangle$ screw dislocations in the equiatomic refractory concentrated alloy (RCA) $\text{NbMo}$ at room temperature under applied deformation rates ranging from $5.0\times10^{7}$ to $1.0\times10^{4}\ \mathrm{s^{-1}}$. For direct comparison with simpler systems, additional simulations are performed for pure Nb, pure Mo, and dilute Mo--Nb alloys under similar loading conditions. All dislocation configurations presented in this study were visualized using the OVITO software package~\cite{stukowski2010visualization}. Dislocation lines were identified and rendered using the Dislocation Extraction Algorithm (DXA) in all subsequent figures~\cite{stukowski2012automated}.

To simulate screw-dislocation glide, a simulation box with cross-sectional dimensions of $10~\mathrm{nm} \times 10~\mathrm{nm}$ was constructed, as shown in Fig.~\ref{fig:MethodHyperdynamics}(b). The simulation cell was populated with the desired atomic species and oriented such that the effective glide direction was $X \parallel [12\bar{1}]$, the glide-plane normal was $Y \parallel [\bar{1}0\bar{1}]$, and the dislocation line direction was $Z \parallel [\bar{1}11]$. After constructing the supercell, a screw dislocation was introduced by imposing a linear displacement field $u_z = b x / l_x$ on all atoms in the upper half of the simulation domain~\cite{Maresca2020TheoryScrewDislocation}. Periodic boundary conditions were applied along the $X$ and $Z$ directions, while traction-free boundary conditions were imposed along the $Y$ direction.

Following dislocation insertion, the atomic positions were relaxed via static energy minimization using the conjugate-gradient (CG) algorithm, with a force tolerance of $\leq 10^{-10}\ \mathrm{eV/\text{\AA}}$ for each atom. Atomic velocities corresponding to a target temperature of $300~\mathrm{K}$ were then assigned, and the system was dynamically equilibrated for $50{,}000$ MD steps using an NPT ensemble with a timestep of $\Delta t = 0.002~\mathrm{ps}$. At the end of equilibration, the residual stresses satisfied $\sigma_{XX}, \sigma_{ZZ} < 5~\mathrm{MPa}$ and $\sigma_{XZ} < 1~\mathrm{MPa}$.

Subsequently, shear deformation was applied by imposing displacement-controlled loading on the top and bottom two atomic layers parallel to the $(\bar{1}0\bar{1})$ plane. The bottom two layers were held fixed, while the top two layers were displaced at a prescribed rate according to
\begin{equation}
\label{eq:AppliedDisplacement}
\dot{u}_z = \dot{\gamma}^{\mathrm{app}}_{yz}\, l_y ,
\end{equation}
where $\dot{\gamma}^{\mathrm{app}}_{yz}$ is the target applied shear strain rate. The incremental displacement at each MD step is given by
$\Delta u_z(t+\Delta t_{\mathrm{MD}}) = \dot{u}_z \Delta t_{\mathrm{phy}}$,
where $\Delta t_{\mathrm{MD}}$ is the MD timestep and $\Delta t_{\mathrm{phy}}$ is the corresponding physical time increment obtained from SBHD. These two timescales are related through Eq.~\eqref{eq:HD_boost}.

Interatomic interactions in the Nb--Mo system were modeled using the angular-dependent potential (ADP) developed by Starikov \emph{et al.}~\cite{Starikov2024AngulardependentInteratomicPotential}. This potential was chosen because it was specifically parameterized to reproduce density-functional-theory (DFT)-level accuracy for key properties relevant to plastic deformation, including vacancy energetics, screw-dislocation cores, and planar defect structures. The ADP has previously been applied to studies of screw-dislocation mobility~\cite{Starikov2025DislocationMobilityFunction}, shock-induced plasticity~\cite{Bryukhanov2025AtomisticSimulationShock}, and grain-boundary structural transitions~\cite{Geiger2024FrustratedMetastabletoequilibriumGrain} in refractory metals and alloys. An alternative interatomic potential for the Nb--Mo system is the SNAP machine-learning potential developed by Li \emph{et al.}~\cite{Li2020ComplexStrengtheningMechanisms}. Although machine-learning potentials can reproduce a broad range of properties with high accuracy, the SNAP potential in Ref.~\cite{Li2020ComplexStrengtheningMechanisms} exhibits a significant deviation from DFT predictions for the Peierls barrier associated with screw-dislocation glide in both Mo and Nb (see Fig.~8 of Ref.~\cite{Starikov2024AngulardependentInteratomicPotential}), motivating the use of the ADP in the present work.

To capture the influence of dislocation line length and elastic line-tension effects discussed in the Introduction, three simulation cell sizes corresponding to dislocation line lengths of approximately $15~\mathrm{nm}$, $20~\mathrm{nm}$, and $50~\mathrm{nm}$ were examined. These lengths span the transition from short periodic dislocation segments, where image interactions may artificially constrain kink formation and migration, to longer segments where configurational roughening and realistic line-tension effects emerge. The rationale for these choices and their consequences for screw-dislocation mobility, kink evolution, and cross-kink interactions are analyzed in detail in the following sections. In addition, two characteristic stresses are extracted from the stress--strain response: (i) the onset resolved shear stress (ORSS), defined as the stress at which the first local deviation from linear elasticity occurs due to activation of a dislocation segment; and (ii) the critical resolved shear stress (CRSS), corresponding to the maximum stress required to depin and mobilize the entire screw-dislocation line.

\subsection{Determination of Parameters Used in Hyperdynamics}
\label{sec:Model_Parameters}

The SBHD method requires specification of several key parameters, including the critical strain threshold $q_{\mathrm{c}}$, the maximum boost amplitude $V_{\mathrm{max}}$, and the number of atoms $N_{\mathrm{b}}$ subjected to the boost potential. These parameters must be chosen carefully to ensure sufficient acceleration while preserving the validity of the TST assumptions underlying hyperdynamics.

\paragraph{Determination of the critical strain threshold $q_{\mathrm{c}}$.}
The threshold parameter $q_{\mathrm{c}}$ defines the onset of a transition event and enforces the condition that the boost potential vanishes as the system approaches a dividing surface between adjacent potential-energy basins. Because local atomic strain contains contributions from both thermal vibrations and structural rearrangements, the optimal value of $q_{\mathrm{c}}$ is inherently temperature dependent. In the present study, $q_{\mathrm{c}}$ is determined using an iterative procedure based on short conventional MD simulations of systems containing a preexisting screw dislocation.

Figures~\ref{fig:MethodHyperdynamics}(c) and (d) show the evolution of the maximum von Mises atomic strain, $\eta^{\mathrm{Mises}}_{\mathrm{max}}$, during shear deformation at $T=300~\mathrm{K}$ for pure Nb and equiatomic NbMo, respectively. In pure Nb (Fig.~\ref{fig:MethodHyperdynamics}(c)), small fluctuations in $\eta^{\mathrm{Mises}}_{\mathrm{max}}$ occur at early stages of deformation (e.g., at $\gamma_{yz}\approx0.21\%$, labeled ``1''), corresponding to unsuccessful attempts at double-kink nucleation. A pronounced and irreversible increase in $\eta^{\mathrm{Mises}}_{\mathrm{max}}$ is observed at $\gamma_{yz}\approx0.38\%$ (labeled ``2''), marking the occurrence of a rare event, namely double-kink nucleation.

For the NbMo alloy (Fig.~\ref{fig:MethodHyperdynamics}(d)), multiple precursor fluctuations (labeled ``1--3'') are observed prior to the transition, reflecting the increased frequency of kink-nucleation attempts induced by local lattice distortions arising from chemical heterogeneity. A clear transition event occurs at $\gamma_{yz}\approx0.40\%$ (labeled ``4''), accompanied by a large increase in $\eta^{\mathrm{Mises}}_{\mathrm{max}}$. Based on these observations, a value of $\eta^{\mathrm{Mises}}_{\mathrm{max}}=0.1$ is identified as a conservative and robust choice for $q_{\mathrm{c}}$ for both pure and alloy systems at $300~\mathrm{K}$.

\paragraph{Determination of the boost amplitude $V_{\mathrm{max}}$.}
The parameter $V_{\mathrm{max}}$ controls the maximum achievable boost and thus the degree of temporal acceleration. It must be sufficiently large to accelerate rare events, yet small enough to avoid the creation of artificial local minima that would violate TST assumptions. Combining Eqs.~\eqref{eq:StrainBoost_PE}, \eqref{eq:StrainBoost_PEi}, and \eqref{eq:StrainBoost_StoppingFunction}, the total boost potential can be written as
\begin{equation}
\label{eq:FullBP}
\Delta V(\mathbf{r}) =
\begin{cases}
\displaystyle \frac{S}{N_{\mathrm{b}}} \sum_i \left[ 1 - \left(\frac{\eta^\mathrm{Mises}_i}{q_{\mathrm{c}}}\right)^2 \right], & \eta^\mathrm{Mises}_{\mathrm{max}} < q_{\mathrm{c}}, \\
0, & \eta^\mathrm{Mises}_{\mathrm{max}} \geq q_{\mathrm{c}},
\end{cases}
\end{equation}
where the quantity
\begin{equation}
\label{eq:StrengthOfBias}
S = V_{\mathrm{max}} \left[ 1 - \left(\frac{\eta^{\mathrm{Mises}}_{\mathrm{max}}}{q_{\mathrm{c}}}\right)^2 \right]
\end{equation}
defines the effective \emph{strength of the bias}.

At the onset of a transition, $S$ should be comparable to the relevant activation barrier. For the ADP potential used here, the Peierls barrier for $\langle111\rangle$ screw-dislocation glide in Nb and Mo is approximately $0.045~\mathrm{eV}$ per Burgers vector (see Fig.~8 of Ref.~\cite{Starikov2024AngulardependentInteratomicPotential}). Accordingly, $S$ is estimated as $S \approx 0.045~\mathrm{eV/atom}$. Since the system frequently samples configurations with $\eta^{\mathrm{Mises}}_{\mathrm{max}} \approx 0.9 q_{\mathrm{c}}$ at $300~\mathrm{K}$, $V_{\mathrm{max}}$ is chosen as $0.21~\mathrm{eV}$. This conservative choice ensures that the boost potential remains consistent with TST, at the expense of some reduction in the achievable acceleration factor.

\paragraph{Selection of boosted atoms $N_{\mathrm{b}}$.}
The number of atoms included in the boost potential directly affects both computational efficiency and accuracy. Ideally, only atoms participating in the transition process should be boosted; however, these atoms cannot be identified \emph{a priori}. Including all atoms in the simulation domain would be overly conservative and computationally prohibitive. In the present study, an intermediate strategy is adopted: only atoms located within the central region of the simulation cell, encompassing the dislocation core and its expected slip path (as shown in Fig.~\ref{fig:MethodHyperdynamics}(b)), are subjected to the boost potential. This choice balances computational efficiency with reliable acceleration of the relevant dislocation-mediated transition events.

\section{Results}
\label{sec:Results}

\subsection{Dislocation line length 15 nm}\label{sec:Results_15nm}

\begin{figure}[!htb]
     \centering
     {\includegraphics[width=1.0\textwidth]{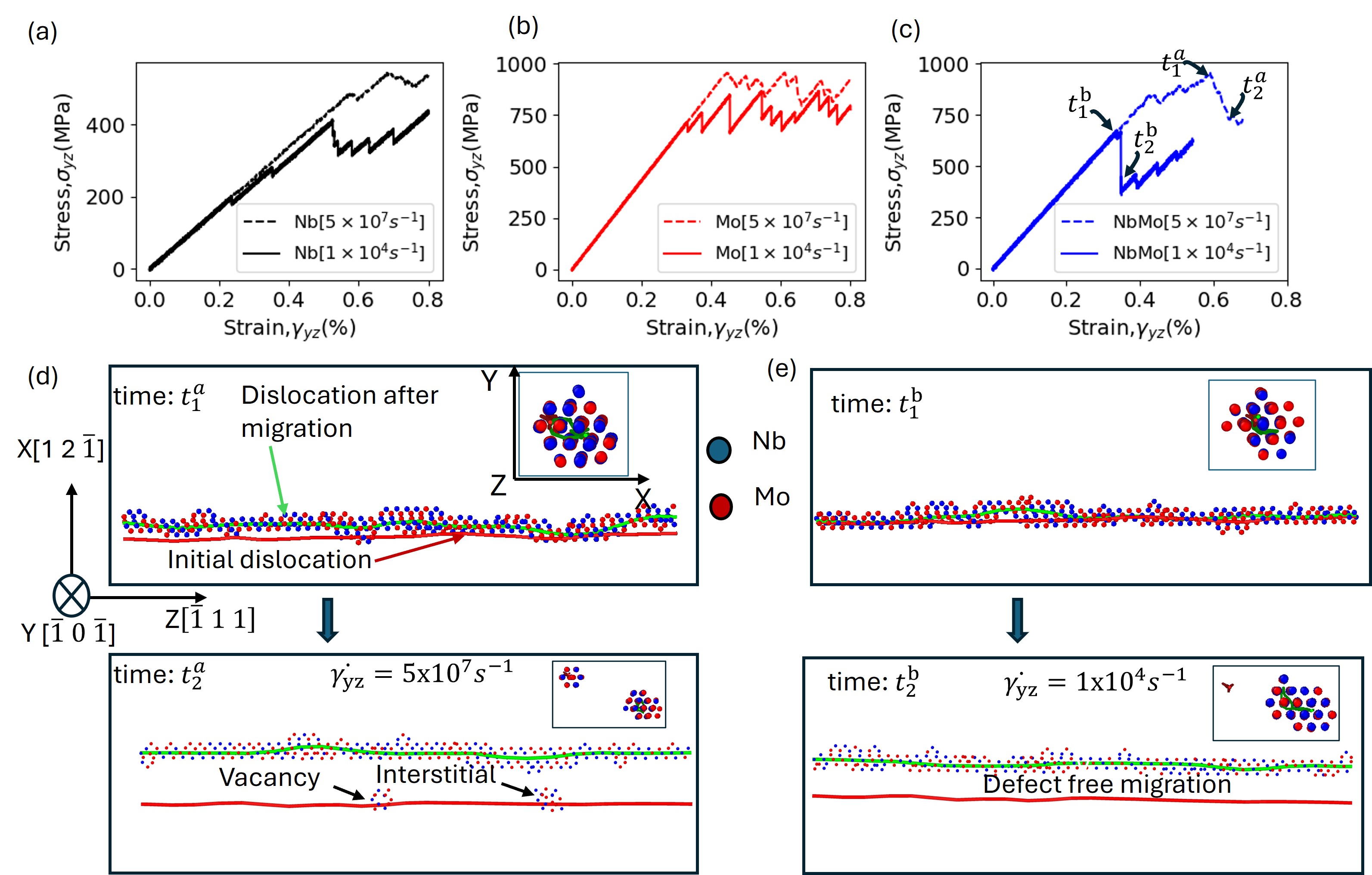}}\label{fig:ResultsCombined_15nm}
\caption {Simulation results for $\langle111\rangle$ screw dislocations with a line length of 15~nm. Shear stress–strain ($\sigma_{yz}$–$\gamma_{yz}$) responses at different applied strain rates for (a) pure Nb, (b) pure Mo, and (c) equiatomic NbMo alloy. (d) Dislocation configurations in NbMo immediately before and after $\sigma_{yz}$ drops at a high strain rate of $5.0\times10^{7}\ \mathrm{s^{-1}}$, illustrating defect generation during cross-kink depinning. (e) Corresponding dislocation configurations in NbMo at a low strain rate of $1.0\times10^{4}\ \mathrm{s^{-1}}$, shown immediately before and after $\sigma_{yz}$ drops.}.
\label{fig:Results15nm}
\end{figure}

A pre-existing $\langle111\rangle$ screw dislocation was introduced into a simulation supercell with dimensions of $10~\mathrm{nm} \times 10~\mathrm{nm} \times 15~\mathrm{nm}$ for pure Nb, pure Mo, and the NbMo alloy. In the pure Nb and Mo systems, the relaxed screw dislocation remains straight and aligned along a single Peierls potential valley. Upon application of increasing shear stress, the dislocation glides via the conventional double-kink nucleation and kink migration mechanism characteristic of BCC metals. Figures~\ref{fig:Results15nm}(a) and (b) show the macroscopic volume-averaged shear stress--strain responses ($\sigma_\mathrm{yz}$--$\gamma_\mathrm{yz}$) for pure Nb and pure Mo at both high ($\dot{\gamma}_\mathrm{yz} = 5.0\times10^{7}\ \mathrm{s^{-1}}$) and low ($\dot{\gamma}_\mathrm{yz} = 1.0\times10^{4}\ \mathrm{s^{-1}}$) strain rates, respectively. Even in these pure BCC metals, the CRSS decreases markedly as the strain rate is reduced, reflecting the thermally activated nature of screw-dislocation motion. At lower strain rates, the increased time available for thermal activation facilitates double-kink nucleation and migration, thereby reducing the applied stress required to sustain dislocation glide.

In the concentrated NbMo alloy, the relaxed dislocation configuration obtained from MD relaxation does not exhibit any cross-kink features. The simulation cell was subsequently subjected to simple shear deformation at constant shear strain rates of $\dot{\gamma}_\mathrm{yz} = 5.0\times10^{7}\,\mathrm{s^{-1}}$ and $\dot{\gamma}_\mathrm{yz} = 1.0\times10^{4}\,\mathrm{s^{-1}}$. Figure~\ref{fig:Results15nm}(c) presents the corresponding shear stress--strain responses under these two strain rates. At the higher strain rate, the CRSS of the NbMo alloy is comparable to that of pure Mo. In contrast, when deformed at the lower strain rate, the CRSS of the NbMo alloy falls between those of pure Nb and pure Mo. To elucidate the origin of this pronounced strain-rate dependence in CRSS, we examine the atomistic dislocation configurations during glide to identify the underlying migration mechanisms.

Figures~\ref{fig:Results15nm}(d) and (e) show representative dislocation configurations in the NbMo alloy immediately before and after the macroscopic stress drop, corresponding to the stage at which each segment of the dislocation line has advanced by at least one Peierls valley. Figure~\ref{fig:Results15nm}(d) corresponds to the configurations at times $t_1^a$ and $t_2^a$ for the $\dot{\gamma}_\mathrm{yz} = 5.0\times10^{7}\ \mathrm{s^{-1}}$ case indicated in Fig.~\ref{fig:Results15nm}(c). Under this high strain rate, deformation leads to the formation of vacancy and interstitial clusters (debris) along the dislocation line near its initial position, highlighted by the red reference line. This behavior indicates that dislocation glide proceeds through cross-kink depinning via point-defect generation, consistent with the mechanism illustrated in Fig.~\ref{fig:GraphicalAbstract}(b). Although the stress-free equilibrium configuration does not contain cross-kinks, the deformation process induces their formation because local segments of the dislocation line glide onto different slip planes. These cross-kinks act as strong pinning points and consequently increase the CRSS. In addition, comparison of the inserted subfigures viewed along the dislocation-line ($Z$) direction at times $t_1^a$ and $t_2^a$ reveals that the dislocation trajectory deviates significantly from the original $(\bar{1}0\bar{1})$ plane (normal to the $Y$ axis), indicating pronounced non-Schmid effects~\cite{Cai2004DislocationCoreEffects,dezerald2016plastic}.

In contrast, Fig.~\ref{fig:Results15nm}(e) shows the configurations at times $t_1^b$ and $t_2^b$ for the lower strain-rate case ($\dot{\gamma}_\mathrm{yz} = 1.0\times10^{4}\ \mathrm{s^{-1}}$). At this lower strain rate, no point-defect debris is observed along the dislocation line, either at its initial position or elsewhere. This behavior suggests two possible scenarios: (i) the dislocation undergoes successive in-plane and lateral kink nucleation events, with the entire dislocation core ultimately gliding along the primary slip plane and thereby avoiding the formation of stable cross-kinks; or (ii) transient cross-kinks do form during glide, but the system accesses lower-energy depinning pathways that do not involve permanent point-defect generation. In either case, the smoother dislocation motion at the lower strain rate leads to a reduced CRSS compared with that observed under high-rate deformation. In addition, comparison of the inserted subfigures viewed along the dislocation-line ($Z$) direction at times $t_1^b$ and $t_2^b$ reveals that, although the dislocation trajectory still deviates from the original $(\bar{1}0\bar{1})$ plane, the extent of deviation is smaller than that observed in Figures~\ref{fig:Results15nm}(d), indicating that the strain rate also influences the non-Schmid behavior.

\subsection{Dislocation line length 20 nm}\label{sec:Results_20nm}

\begin{figure}[!htb]
     \centering
     {\includegraphics[width=1.0\textwidth]{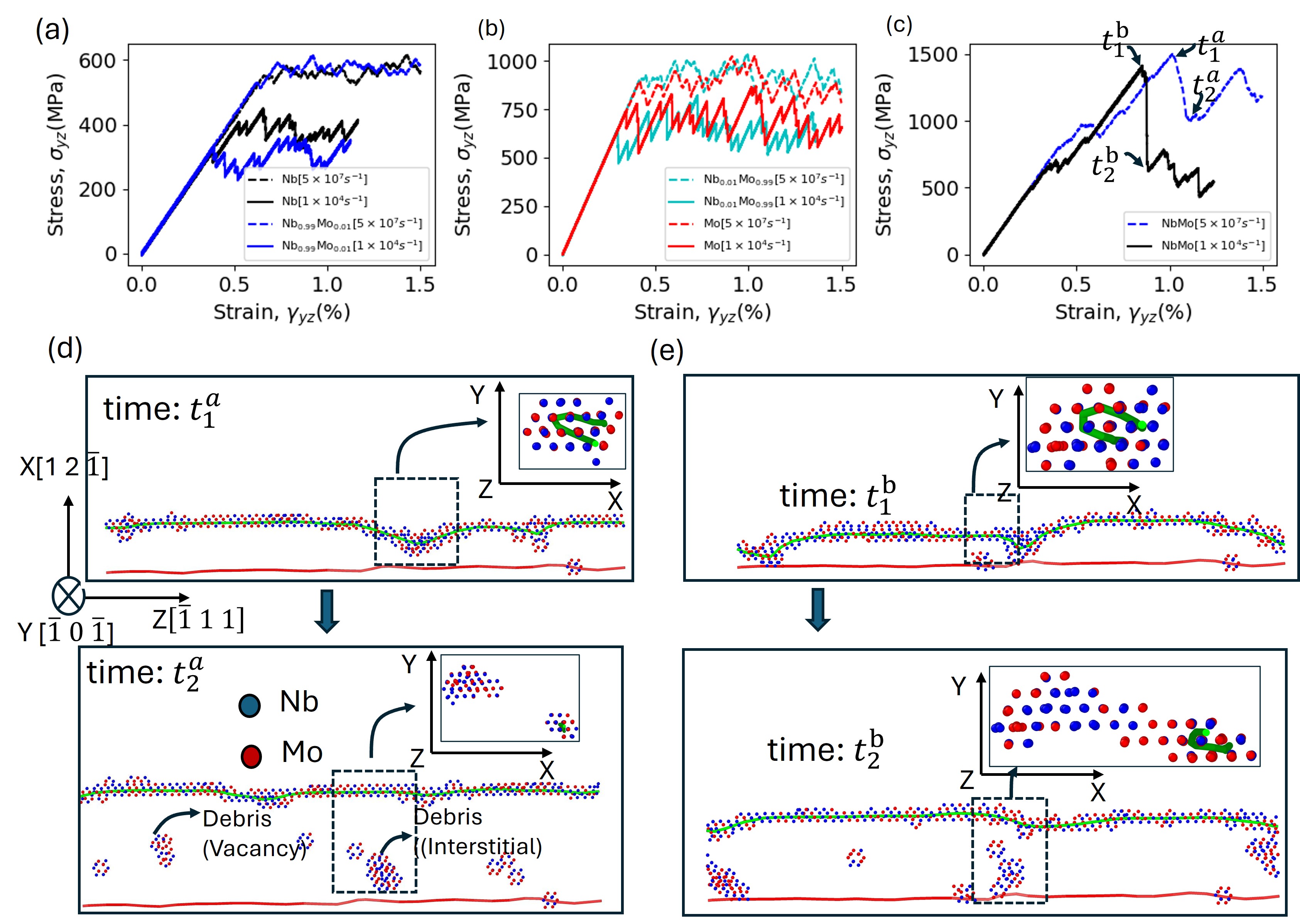}}\label{fig:ResultsCombined_20nm}
\caption {Simulation results for screw dislocations with a line length of 20 nm. Shear stress–strain ($\sigma_{yz}$–$\gamma_{yz}$) responses at different applied strain rates for (a) pure Nb, (b) pure Mo, and (c) the NbMo alloy. (d) Dislocation configurations in the NbMo alloy immediately before and after $\sigma_{yz}$ drops under a high strain rate of $5.0\times10^{7}\ \mathrm{s^{-1}}$. (e) Dislocation configurations in the NbMo alloy immediately before and after $\sigma_{yz}$ drops under a low strain rate of $1.0\times10^{4}\ \mathrm{s^{-1}}$.}
\label{fig:Results20nm}
\end{figure}

Next, the dislocation line length was increased to $20~\mathrm{nm}$ by extending the supercell dimension along the screw-dislocation line direction in order to more accurately capture elastic line-tension effects under different shear strain rates. Figures~\ref{fig:Results20nm}(a) and (b) present the shear stress--strain responses ($\sigma_{yz}$--$\gamma_{yz}$) for pure Nb, pure Mo, and two dilute alloys, $\text{Nb}_{0.99}\text{Mo}_{0.01}$ and $\text{Nb}_{0.01}\text{Mo}_{0.99}$. As expected, all systems exhibit a pronounced reduction in the CRSS as the strain rate decreases, consistent with the thermally activated nature of screw-dislocation motion. This trend is particularly evident for pure Nb and $\text{Nb}_{0.99}\text{Mo}_{0.01}$ in Fig.~\ref{fig:Results20nm}(a). More interestingly, because the solute concentrations in $\text{Nb}_{0.99}\text{Mo}_{0.01}$ and $\text{Nb}_{0.01}\text{Mo}_{0.99}$ are low, solute atoms are expected to produce a mild \emph{solute softening} effect~\cite{trinkle2005chemistry,hu2017solute}: they locally assist kink nucleation without significantly impeding kink migration. This effect becomes increasingly pronounced at lower strain rates. Specifically, when comparing the low-rate ($\dot{\gamma}_\mathrm{yz} = 1.0\times10^{4}\,\mathrm{s^{-1}}$) results with the high-rate ($\dot{\gamma}_\mathrm{yz} = 5.0\times10^{7}\,\mathrm{s^{-1}}$) cases, both the CRSS and the average flow stress of the dilute alloys are noticeably lower than those of their corresponding pure metals (Nb and Mo), as shown by the solid curves in Figs.~\ref{fig:Results20nm}(a) and (b). This observation confirms that solute softening is more effective at slower deformation rates, where kink nucleation processes are more strongly thermally activated.

Figure~\ref{fig:Results20nm}(c) shows the $\sigma_{yz}$--$\gamma_{yz}$ responses of the NbMo alloy at the same two strain rates for a dislocation line length of $20~\mathrm{nm}$. In contrast to the shorter-line-length case shown in Fig.~\ref{fig:Results15nm}(c), the CRSS for the $\dot{\gamma}_{yz} = 1.0\times10^{4}\ \mathrm{s^{-1}}$ case does not decrease significantly relative to that for $\dot{\gamma}_{yz} = 5.0\times10^{7}\,\mathrm{s^{-1}}$, although the post-yield stress--strain behaviors of the two cases differ substantially. Inspection of the relaxed dislocation configurations reveals that, for the longer dislocation line, cross-kink structures are already present prior to deformation and act as strong local pinning points in both strain-rate regimes. Consequently, the CRSS of NbMo for the $20~\mathrm{nm}$ cases ($\sim1.5~\mathrm{GPa}$) is significantly higher than that for the $15~\mathrm{nm}$ cases in Fig.~\ref{fig:Results15nm}(c) (less than $1.0~\mathrm{GPa}$), highlighting the important role of dislocation line length.

Figure~\ref{fig:Results20nm}(d) shows the dislocation configurations at times $t_1^{a}$ and $t_2^{a}$ for the $\dot{\gamma}_\mathrm{yz} = 5.0\times10^{7}\,\mathrm{s^{-1}}$ case indicated in Fig.~\ref{fig:Results20nm}(c). Under the high strain rate, the dislocation undergoes extensive cross-slip among multiple $\{110\}$ planes, causing the cross-kink formation and the bowing of the dislocation line on different slip planes before depinning occurs. The subsequent annihilation of these cross-kinks leaves behind large amounts of debris in the form of vacancy and interstitial clusters. Similar features are observed in Fig.~\ref{fig:Results20nm}(e), which corresponds to the configurations at times $t_1^b$ and $t_2^b$ for the $\dot{\gamma}_\mathrm{yz} = 1.0\times10^{4}\,\mathrm{s^{-1}}$ case in Fig.~\ref{fig:Results20nm}(c). However, notable differences emerge between the two strain-rate regimes.

As shown in Fig.~\ref{fig:Results20nm}(d), the dislocation configuration at time $t_2^a$ in the high–strain-rate case exhibits a glide trajectory that, when viewed along the dislocation-line ($Z$) direction in the inserted subfigures, deviates significantly from the original $\{110\}$ planes. This behavior reflects the averaged contribution of slip on multiple $\{110\}$ planes and strong non-Schmid effects~\cite{Cai2004DislocationCoreEffects,dezerald2016plastic}. In contrast, at the lower strain rate of $\dot{\gamma}_\mathrm{yz} = 1.0\times10^{4}\ \mathrm{s^{-1}}$, as shown in Fig.~\ref{fig:Results20nm}(e), dislocation glide occurs predominantly on a single $(\bar{1}0\bar{1})$ slip plane. In addition, the amount of debris generated is noticeably smaller, which explains why a lower applied stress is sufficient to sustain dislocation motion after $\sigma_\mathrm{yz}$ reaches the CRSS, as indicated in Fig.~\ref{fig:Results20nm}(c). Together, these results demonstrate that both strain rate and dislocation line length strongly influence cross-slip activity, cross-kink evolution, and depinning mechanisms in the NbMo alloy, underscoring the necessity of further increasing the dislocation line length to more accurately capture the intrinsic mechanisms of screw-dislocation motion.

\subsection{Dislocation line length 50 nm for pure metals and dilute alloys}\label{sec:Results_50nm_a}

\begin{figure}[!htb] 
\centering
    \includegraphics[width=1.0\textwidth]{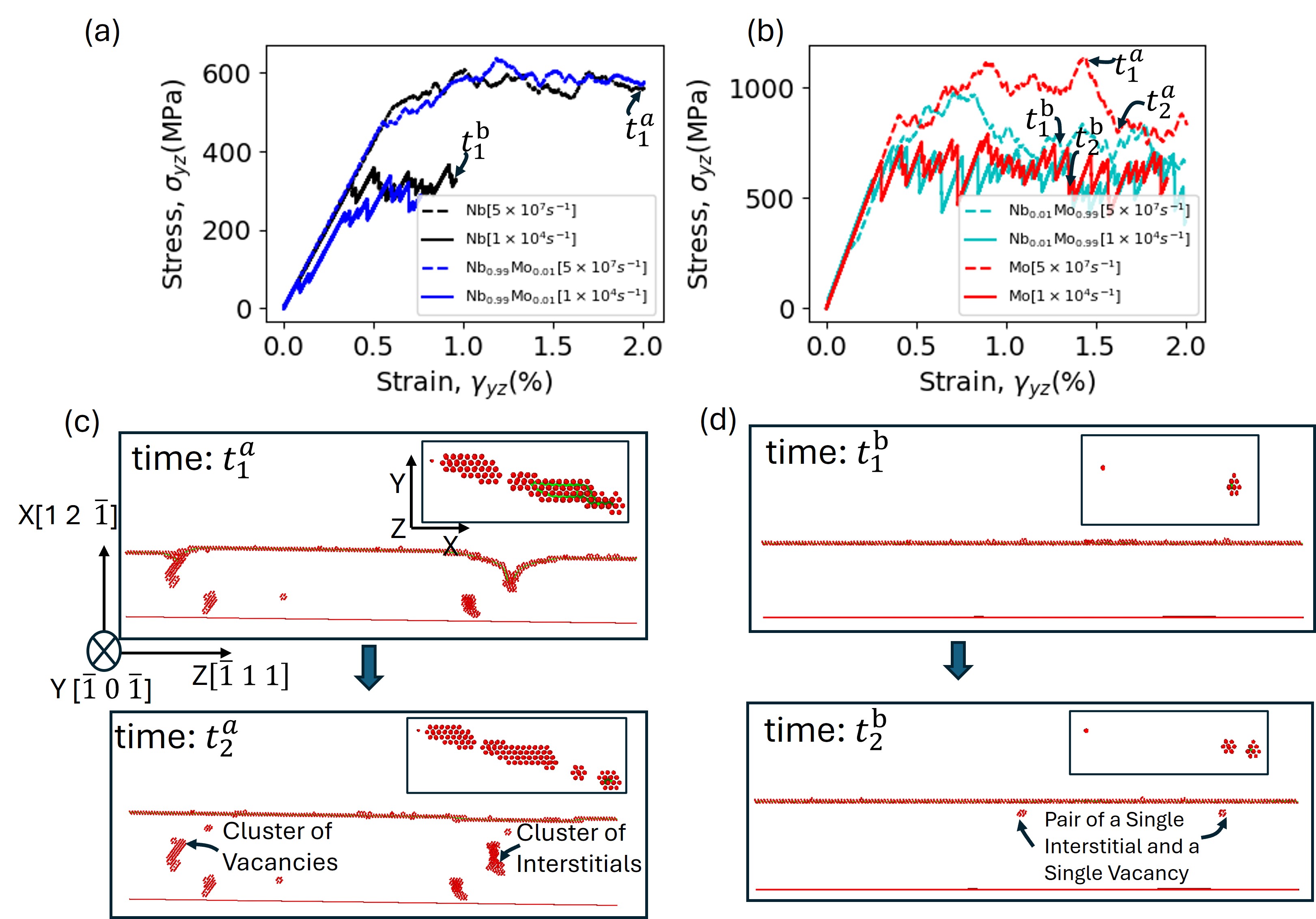}
    \caption{Simulation results for screw dislocations with a line length of 50 nm in pure Nb, pure Mo, and dilute Nb–Mo alloys. Shear stress–strain ($\sigma_{yz}$–$\gamma_{yz}$) responses for (a) pure Nb and $\text{Nb}_{0.99}\text{Mo}_{0.01}$, and (b) pure Mo and $\text{Nb}_{0.01}\text{Mo}_{0.99}$.  Dislocation and defect configurations immediately before and after depinning in pure Mo at strain rates of (c) $\dot{\gamma}_{yz} = 5.0\times10^{7}\ \mathrm{s^{-1}}$ and (d) $\dot{\gamma}_{yz} = 1.0\times10^{4}\ \mathrm{s^{-1}}$.}
    \label{fig:Results_50nm_Nb_and_Mo} 
\end{figure}

Next, to more accurately capture the coupled effects of strain rate and dislocation line length, we extend the dislocation line length to $50~\mathrm{nm}$ in our molecular dynamics simulations and systematically examine screw dislocation glide under different applied strain rates. In addition, to isolate and clarify the role of chemical complexity, we first focus on pure BCC metals (Nb and Mo) and their dilute alloys, $\text{Nb}_{0.99}\text{Mo}_{0.01}$ and $\text{Nb}_{0.01}\text{Mo}_{0.99}$, before analyzing the behavior of concentrated NbMo alloys. Previous atomistic studies have shown that dislocation line lengths on the order of $\sim 50~\mathrm{nm}$ are sufficient to accurately capture kink structures and their elastic interactions in BCC screw dislocations, thereby providing a more reliable description of line-tension and kink–kink interaction effects than shorter simulation cells~\cite{ji2020quantifying}.

Figure~\ref{fig:Results_50nm_Nb_and_Mo}(a) and (b) show the $\sigma_{yz}$–$\gamma_{yz}$ responses for pure Nb, pure Mo, and the corresponding dilute binary alloys $\text{Nb}_{0.99}\text{Mo}_{0.01}$ and $\text{Nb}_{0.01}\text{Mo}_{0.99}$ for a dislocation line length of $50~\mathrm{nm}$. Compared with the $20~\mathrm{nm}$ results shown in Figure~\ref{fig:Results20nm}(a), the strain-rate effects are noticeably stronger when the dislocation line is extended to $50~\mathrm{nm}$. In particular, both the onset resolved shear stress (ORSS) required to initiate local segment motion and the critical resolved shear stress (CRSS) associated with mobilizing the entire screw dislocation exhibit a larger reduction as the strain rate decreases. For example, in pure Nb, the high–strain-rate CRSS remains approximately $600~\mathrm{MPa}$ for both the $20~\mathrm{nm}$ and $50~\mathrm{nm}$ line-length cases, whereas the low–strain-rate CRSS decreases from more than $400~\mathrm{MPa}$ at $20~\mathrm{nm}$ to about $300~\mathrm{MPa}$ at $50~\mathrm{nm}$. A similar trend is observed in pure Mo: the high–strain-rate CRSS exceeds $1000~\mathrm{MPa}$ for both line lengths, while the low–strain-rate value decreases from roughly $800~\mathrm{MPa}$ in the $20~\mathrm{nm}$ case to about $750~\mathrm{MPa}$ at $50~\mathrm{nm}$. In addition, the $50~\mathrm{nm}$ simulations reveal a pronounced solute-softening effect in the dilute alloys. The ORSS required to initiate local dislocation-segment motion is substantially lower in the dilute alloys at $50~\mathrm{nm}$ compared with the $20~\mathrm{nm}$ results, as seen by comparing the curves for $\text{Nb}_{0.99}\text{Mo}_{0.01}$ and $\text{Nb}_{0.01}\text{Mo}_{0.99}$ in Figure~\ref{fig:Results_50nm_Nb_and_Mo}(a) and (b) with those in Figure~\ref{fig:Results20nm}(a). These observations indicate an enhanced influence of elastic line tension on kink nucleation and migration at larger dislocation line lengths.

More interestingly, as shown in Figures~\ref{fig:Results_50nm_Nb_and_Mo}(c) and (d), the strain rate strongly influences screw-dislocation glide mechanisms even in pure Mo, including the preferred slip planes and the manner in which local pinning points are overcome. In the high–strain-rate case (Figure~\ref{fig:Results_50nm_Nb_and_Mo}(c)), the screw dislocation in pure Mo repeatedly forms cross-kinks as local segments glide on different $\{110\}$-type planes. These cross-kinks act as strong pinning points, visible as curved segments of the dislocation line at time $t_{1}^{a}$, corresponding to the CRSS in Figure~\ref{fig:Results_50nm_Nb_and_Mo}(b). Upon depinning (e.g., at time $t_{2}^{a}$, immediately after the stress drop), vacancy and interstitial clusters are generated, indicating that depinning proceeds through cross-kink formation and drag mechanisms, as illustrated in Fig.~\ref{fig:GraphicalAbstract}(b). The inset views along the dislocation-line direction ($Z \parallel [\bar{1}11]$) further show that the trajectory oscillates among multiple slip planes, leading to substantial deviation from the $(\bar{1}0\bar{1})$ plane.

In contrast, at the low strain rate for pure Mo (Figure~\ref{fig:Results_50nm_Nb_and_Mo}(d)), corresponding to times $t_{1}^{b}$ and $t_{2}^{b}$ in Figure~\ref{fig:Results_50nm_Nb_and_Mo}(b), the dislocation remains relatively straight and produces minimal debris (only a single vacancy and interstitial at $t_{2}^{b}$). The inset trajectory indicates that glide is largely confined to the $(\bar{1}0\bar{1})$ plane, which carries the highest resolved shear stress, with only minor deviations. These results demonstrate that, although cross-kinks can form even in pure or dilute systems, the depinning pathways depend sensitively on strain rate and elastic line-tension effects, the latter being strongly influenced by the total dislocation line length in periodic supercell simulations.

\begin{figure}[!htb] 
\centering
    \includegraphics[width=1.0\textwidth]{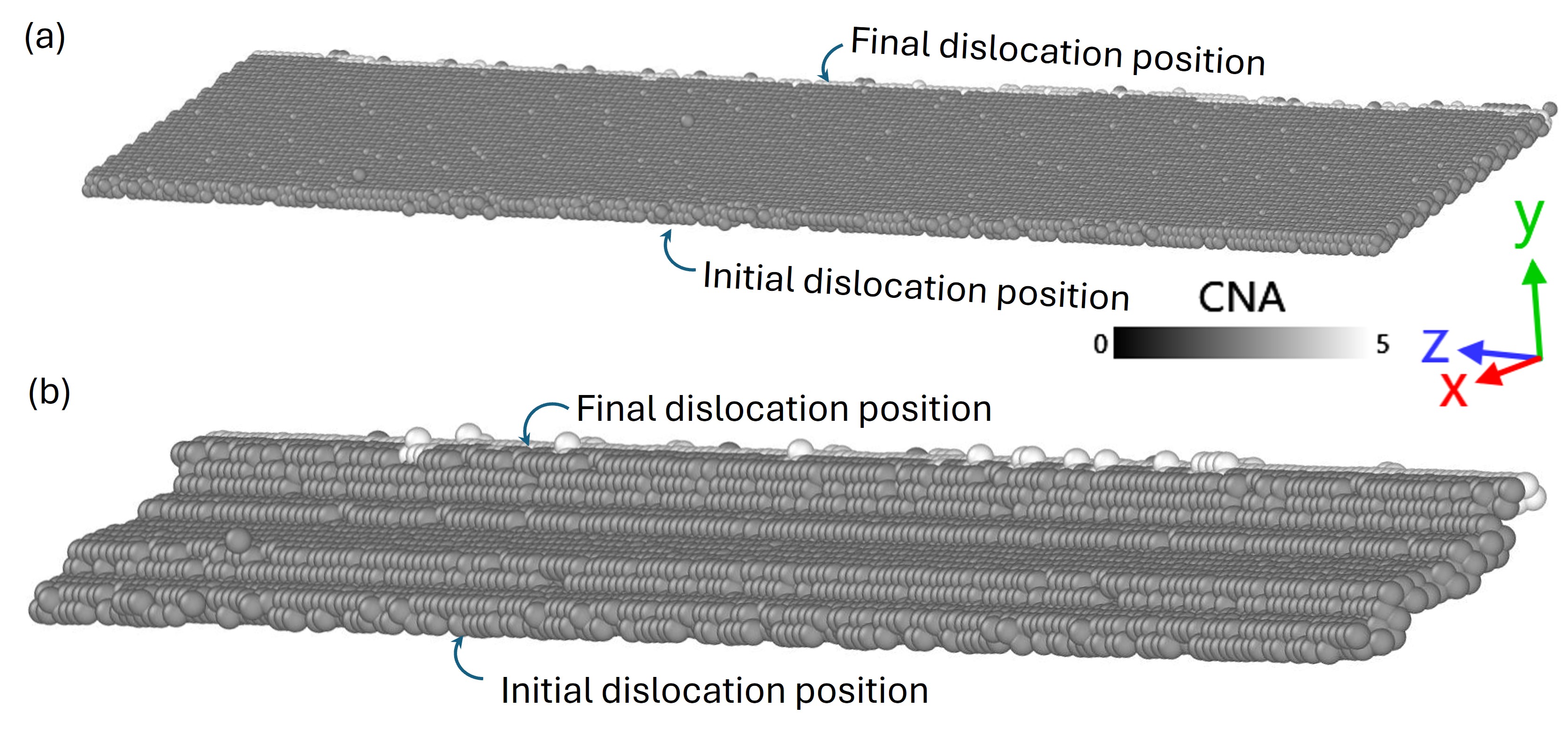}
    \caption{3D dislocation slip trajectories in pure Nb, visualized by displaying only atoms whose $zy$ component of the deformation gradient, $F_{zy}$, exceeds a critical threshold value of 0.2. The selected atoms are colored according to their common neighbor analysis (CNA) values~\cite{honeycutt1987molecular}, scaled between 0 and 5. (a) Slip trajectory constructed from the MD configuration at time $t_1^a$ in Fig.~\ref{fig:Results_50nm_Nb_and_Mo}(a) for pure Nb under a strain rate of $\dot{\gamma}_{yz} = 5.0\times10^{7}\ \mathrm{s^{-1}}$. (b) Slip trajectory constructed from the MD configuration at time $t_1^b$ in Fig.~\ref{fig:Results_50nm_Nb_and_Mo}(a) for pure Nb under a strain rate of $\dot{\gamma}_{yz} = 1.0\times10^{4}\ \mathrm{s^{-1}}$.}
    \label{fig:SlipPath_50nm_Nb} 
\end{figure}

To accurately elucidate the depinning mechanisms discussed above, we plot the 3D screw-dislocation trajectories by selecting atoms whose $zy$ component of the deformation gradient, $F_{zy}$, exceeds a critical threshold value of 0.2. The remaining atoms are then colored according to their common neighbor analysis (CNA) values~\cite{honeycutt1987molecular}, scaled between 0 and 5, such that dislocation core structures and other defect structures generated during dislocation glide are visually distinguished from atoms undergoing simple glide without defect formation. 
As shown in Fig.~\ref{fig:SlipPath_50nm_Nb}(a), when this threshold is applied to the pure Nb system with a $50~\mathrm{nm}$ dislocation line length deformed at a high strain rate ($\dot{\gamma}_{yz} = 5.0\times10^{7}\ \mathrm{s^{-1}}$), and the dislocation glide occurs strictly on a single $(\bar{1}0\bar{1})$ slip plane, only atoms immediately above and below the slip plane are selected. This indicates that the Nb screw dislocation undergoes fully planar glide without any cross-slip events or cross-kink defect formation. In contrast, as shown in Fig.~\ref{fig:SlipPath_50nm_Nb}(b), applying the same threshold to the pure Nb system with the same dislocation line length but deformed at a lower strain rate ($\dot{\gamma}_{yz} = 1.0\times10^{4}\ \mathrm{s^{-1}}$) reveals pronounced step–terrace configurations. Within each flat terrace, atoms are again selected only above and below a single slip plane, indicating that the screw dislocation still undergoes locally planar glide over short distances. However, each step corresponds to a cross-slip event from one $(\bar{1}0\bar{1})$ plane to an adjacent $(\bar{1}0\bar{1})$ plane. Moreover, the ledges along these steps are nearly straight, suggesting that each cross-slip event is initiated by a single double-kink nucleation on a neighboring $\{110\}$ plane, followed by smooth kink migration without the formation of additional defects. This observation indicates that, in pure Nb, double-kink nucleation on either the primary $(\bar{1}0\bar{1})$ plane or the cross-slip plane is the rate-determining step, whereas kink migration is generally rapid under these conditions.

\begin{figure}[!htb] 
\centering
    \includegraphics[width=1.0\textwidth]{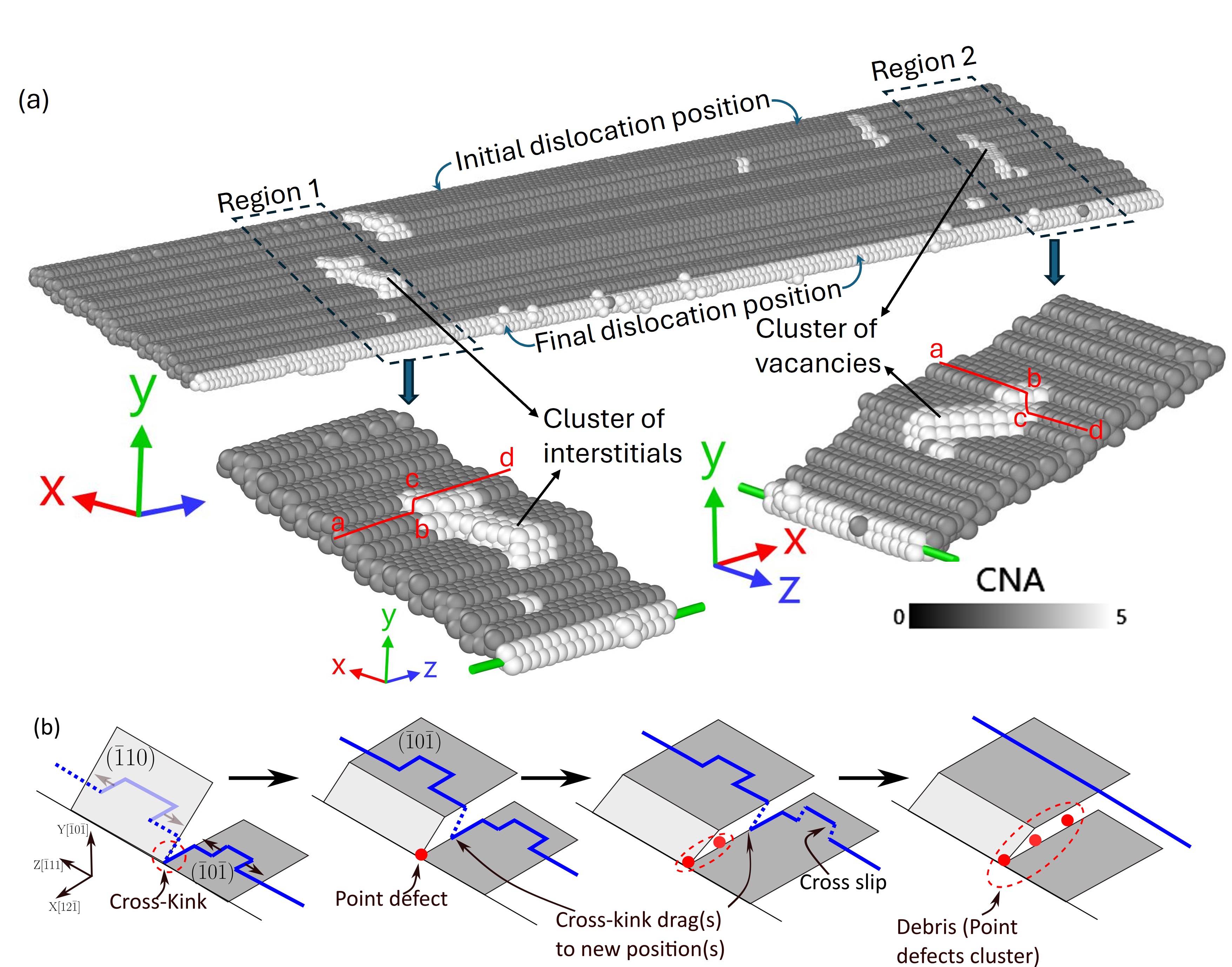}
    \caption{High-strain-rate depinning mechanism in pure Mo illustrating cross-kink depinning via defect formation. (a) 3D dislocation slip trajectory in pure Mo, visualized using the same method described in Fig.~\ref{fig:SlipPath_50nm_Nb}. The trajectory is constructed from the MD configuration at time $t_2^a$ in Fig.~\ref{fig:Results_50nm_Nb_and_Mo}(b) under a strain rate of $\dot{\gamma}_{yz} = 5.0\times10^{7}\ \mathrm{s^{-1}}$. (b) Schematic illustration of the detailed depinning process of a cross-kink through the formation of vacancy and self-interstitial clusters, consistent with the mechanism shown in Fig.~\ref{fig:GraphicalAbstract}(b).}
    \label{fig:SlipPath_50nm_Mo_CrossKinkDragging} 
\end{figure}

Fig.~\ref{fig:SlipPath_50nm_Mo_CrossKinkDragging}(a) shows the three-dimensional screw-dislocation trajectories of pure Mo with a $50~\mathrm{nm}$ dislocation line length deformed at a high strain rate ($5.0\times10^{7}\ \mathrm{s^{-1}}$), obtained using the same deformation-gradient-based analysis described above. The CNA coloring clearly indicates the formation of two debris regions, labeled Region~1 with a cluster of interstitials and Region~2 with a cluster of vacancies, highlighted by dashed rectangles. For each region, the corresponding enlarged subfigure shows that the debris is generated by adjacent dislocation-line segments gliding on different slip planes over finite distances. This behavior is illustrated by the ledges highlighted by red line segments labeled “a”, “b”, “c”, and “d”. Here, the line segment “ab” lies on a different $(\bar{1}0\bar{1})$ plane from segment “cd”, and the connecting segment “bc” represents a kink along the ledge. This configuration provides clear evidence of cross-kink formation during dislocation glide. In addition, the dislocation line eventually glides back onto a single slip plane, as indicated by the reappearance of a terrace–step configuration with long and nearly straight steps. These observations demonstrate that the dislocation line depins from the cross-kink defects and resumes planar glide following the depinning process.

To elucidate the cross-kink depinning process shown in Fig.~\ref{fig:SlipPath_50nm_Mo_CrossKinkDragging}(a), we provide a detailed illustration of the underlying dislocation motion in Fig.~\ref{fig:SlipPath_50nm_Mo_CrossKinkDragging}(b). The leftmost subfigure corresponds to the configuration immediately after cross-kink formation due to dislocation lines on both the original $(\bar{1}0\bar{1})$ plane  (solid blue lines) and the $(\bar{1}10)$ plane (dashed blue lines), as schematically illustrated in Fig.~\ref{fig:GraphicalAbstract} (a). Starting from each of these dislocation line segments, double kinks nucleate on two different but adjacent $(\bar{1}0\bar{1})$ planes and advance along the $[\bar{1}\,\bar{2}\,1]$ direction on both planes. As a result, the distances between these two dislocation segments and the intersection line of the $(\bar{1}0\bar{1})$ and $(\bar{1}10)$ planes increase, enlarging the cross-kink and raising its total energy. This energy increase drives the cross-kink junction to migrate along the $[\bar{1}\,\bar{2}\,1]$ direction. During this dragging process, a point defect forms at the intersection of the $(\bar{1}0\bar{1})$ and $(\bar{1}10)$ planes, as shown in the middle-left subfigure of Fig.~\ref{fig:SlipPath_50nm_Mo_CrossKinkDragging}(b). Repeated cross-kink dragging through the same mechanism, slip on two different $(\bar{1}0\bar{1})$ planes accompanied by point-defect formation, leads to the accumulation of a cluster of point defects, as illustrated in the middle-right subfigure.

As also indicated in the middle-right subfigure of Fig.~\ref{fig:SlipPath_50nm_Mo_CrossKinkDragging}(b), a cross-slip event subsequently occurs in which the dislocation segment on the original $(\bar{1}0\bar{1})$ plane transfers to an adjacent $(\bar{1}10)$ plane through an intermediate configuration. Subsequent kink (dashed blue line segments) migration on the $(\bar{1}10)$ plane enables the entire dislocation line to move onto the new $(\bar{1}0\bar{1})$ plane, leaving behind a large debris cluster composed of point defects, as shown in the rightmost subfigure of Fig.~\ref{fig:SlipPath_50nm_Mo_CrossKinkDragging}(b). This depinning mechanism explains the origin of the debris observed in Fig.~\ref{fig:Results_50nm_Nb_and_Mo}(c). The key driving factor is the rapid forward migration of dislocation segments on the primary $(\bar{1}0\bar{1})$ slip planes, which generates a strong drag force on the cross-kink and promotes point-defect formation. The results further suggest that the size of the debris cluster is governed by the relative rates of cross-slip events that return the dislocation segments to the same $(\bar{1}0\bar{1})$ plane and the kink nucleation and migration processes on the primary slip plane. Consequently, the competition between these rates determines the dominant depinning mechanism and the associated critical stress.

\begin{figure}[!htbp] 
\centering
    \includegraphics[width=0.9\textwidth]{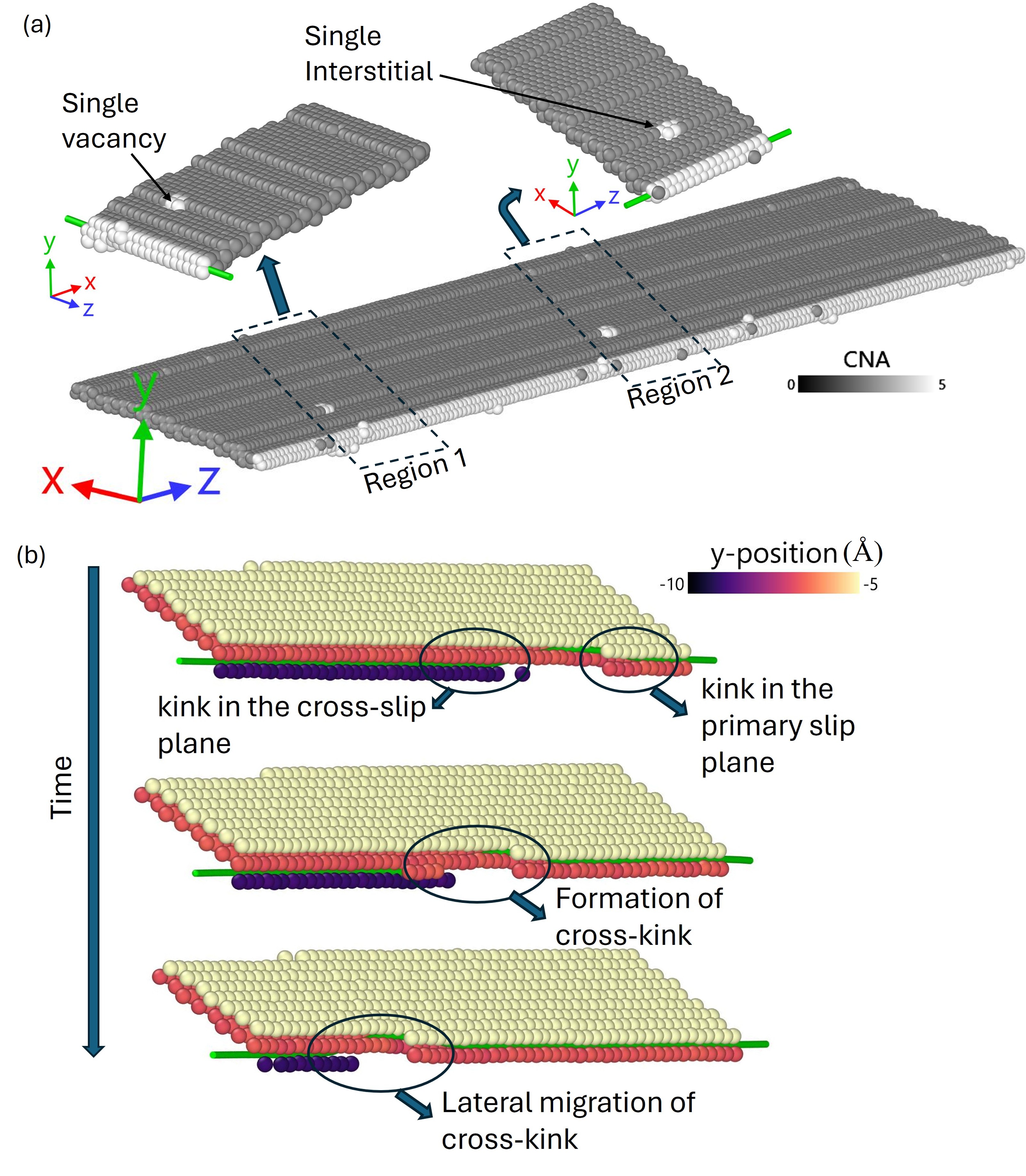}
    \caption{Low-strain-rate depinning mechanism in pure Mo illustrating lateral cross-kink migration without significant defect generation. (a) 3D dislocation slip trajectory in pure Mo, visualized using the same method described in Fig.~\ref{fig:SlipPath_50nm_Nb}. The trajectory is constructed from the MD configuration at time $t_2^b$ in Fig.~\ref{fig:Results_50nm_Nb_and_Mo}(b) under a strain rate of $\dot{\gamma}_{yz} = 1.0\times10^{4}\ \mathrm{s^{-1}}$. (b) Snapshots of the screw-dislocation trajectory in (a) reveal a lateral cross-kink migration mechanism along the dislocation line, consistent with the mechanism illustrated in Fig.~\ref{fig:GraphicalAbstract}(c). In this visualization, the selected atoms are colored according to their relative coordinates along the $y$ axis (normal to the $(\bar{1}0\bar{1})$ plane), with the initial dislocation-line position taken as the reference zero, rather than by their CNA values.}
    \label{fig:SlipPath_50nm_Mo_CrossKinkMigration} 
\end{figure}

To demonstrate how the cross-kink depinning mechanisms discussed above are affected by strain rate, Fig.~\ref{fig:SlipPath_50nm_Mo_CrossKinkMigration} shows the three-dimensional screw-dislocation trajectories of pure Mo with a $50~\mathrm{nm}$ dislocation line length deformed at a low strain rate ($1.0\times10^{4}\ \mathrm{s^{-1}}$). Similar to Fig.~\ref{fig:SlipPath_50nm_Mo_CrossKinkDragging}(a), the trajectories in Fig.~\ref{fig:SlipPath_50nm_Mo_CrossKinkMigration}(a) also exhibit a characteristic terrace–step configuration, arising from frequent cross-slip of the dislocation between adjacent $(\bar{1}0\bar{1})$ planes. In addition, two regions containing debris (Region~1 and Region~2, highlighted by dashed rectangles in Fig.~\ref{fig:SlipPath_50nm_Mo_CrossKinkMigration}(a)) are observed, indicating depinning events associated with cross-kinks. However, there are clear differences in the depinning mechanisms between the high- and low-strain-rate cases. Compared with the large debris clusters composed of multiple vacancies and interstitials in Fig.~\ref{fig:SlipPath_50nm_Mo_CrossKinkDragging}(a), which are generated through the cross-kink dragging mechanism illustrated in Fig.~\ref{fig:SlipPath_50nm_Mo_CrossKinkDragging}(b), the debris observed in Regions~1 and~2 of Fig.~\ref{fig:SlipPath_50nm_Mo_CrossKinkMigration}(a) consists of only a single vacancy or a single interstitial. This difference suggests that, although isolated point defects can still be generated during cross-kink depinning at low strain rates, significant cross-kink dragging does not occur under these conditions. This behavior can be understood by noting that, at low strain rates, dislocation glide on the two different $(\bar{1}0\bar{1})$ planes (as shown in the middle-left subfigure of Fig.~\ref{fig:SlipPath_50nm_Mo_CrossKinkDragging}(b)) proceeds more slowly relative to the cross-slip events that realign the dislocation onto a single $(\bar{1}0\bar{1})$ plane (as illustrated in the middle-right subfigure of Fig.~\ref{fig:SlipPath_50nm_Mo_CrossKinkDragging}(b)). Consequently, after cross-kink formation, the dislocation segments return to the same slip plane more rapidly, reducing the driving force for cross-kink dragging and suppressing the formation of large debris clusters. These results further indicate that the size of the debris cluster is governed by the competition between the rate of cross-slip events that realign the dislocation onto a single $(\bar{1}0\bar{1})$ plane and the rates of kink nucleation and migration on the primary slip plane.

More interestingly, Fig.~\ref{fig:SlipPath_50nm_Mo_CrossKinkMigration}(b) reveals a distinct depinning mechanism for cross-kinks in pure Mo under low strain-rate conditions. In this figure, three-dimensional screw-dislocation trajectories are again visualized by selecting only atoms with the $zy$ component of the deformation gradient, $F_{zy}$, exceeding a critical threshold value of 0.2. Unlike previous figures, however, the selected atoms are colored not according to their CNA values but based on their relative coordinates along the $y$ axis (normal to the $(\bar{1}0\bar{1})$ plane), with the initial dislocation-line position taken as the reference zero. From the top to the bottom panels of Fig.~\ref{fig:SlipPath_50nm_Mo_CrossKinkMigration}(b), the evolution of the dislocation configuration can be clearly observed. Initially, two kinks form: one on the primary $(\bar{1}0\bar{1})$ slip plane and the other on a cross-slip plane, as indicated by their distinct $y$-coordinate values. These kinks then migrate on their respective planes in opposite directions along the dislocation line, leading to the formation of a large cross-kink structure, highlighted by the circle in the middle panel of Fig.~\ref{fig:SlipPath_50nm_Mo_CrossKinkMigration}(b). Subsequently, as shown in the bottom panel, this cross-kink undergoes lateral migration along the dislocation-line direction without generating any additional defects or debris. This depinning process is enabled by the simultaneous migration of the kink on the primary slip plane and the kink on the cross-slip plane in the same direction, which ultimately annihilates the dislocation segment on the cross-slip plane and restores the entire dislocation line onto the primary slip plane. Such lateral cross-kink migration events can occur repeatedly, resulting in the overall terrace–step configuration of the screw-dislocation trajectories in Fig.~\ref{fig:SlipPath_50nm_Mo_CrossKinkMigration}(a), which exhibits nearly straight ledges between adjacent terraces with very few additional defect structures. 

Similar lateral cross-kink migration behavior is likely to occur in pure Nb under the same low strain-rate conditions; however, these events are too rapid to be readily captured in snapshots of the simulation trajectories. As shown in Fig.~\ref{fig:SlipPath_50nm_Nb}(b), this behavior leads to comparable terrace–step configurations with minimal defect accumulation in both Nb and Mo. These observations suggest that, although cross-slip events, initiated by kink nucleation and migration on planes distinct from the primary slip plane, can readily occur in pure BCC metals to form cross-kink pinning points, rapid and cooperative kink migration along the dislocation line at sufficiently low strain rates can efficiently eliminate these cross-kinks without generating additional defects.

%\begin{figure}[H] 
%\centering
%    \includegraphics[width=1.0\textwidth]{Figures/SlipPath_50nm_Diluted_All.jpg}
%    \caption{Dislocation slip path of Diluted alloys.}
%    \label{fig:SlipPath_50nm_Diluted_All} 
%\end{figure}

\begin{figure}[!htb] 
\centering
    \includegraphics[width=1.0\textwidth]{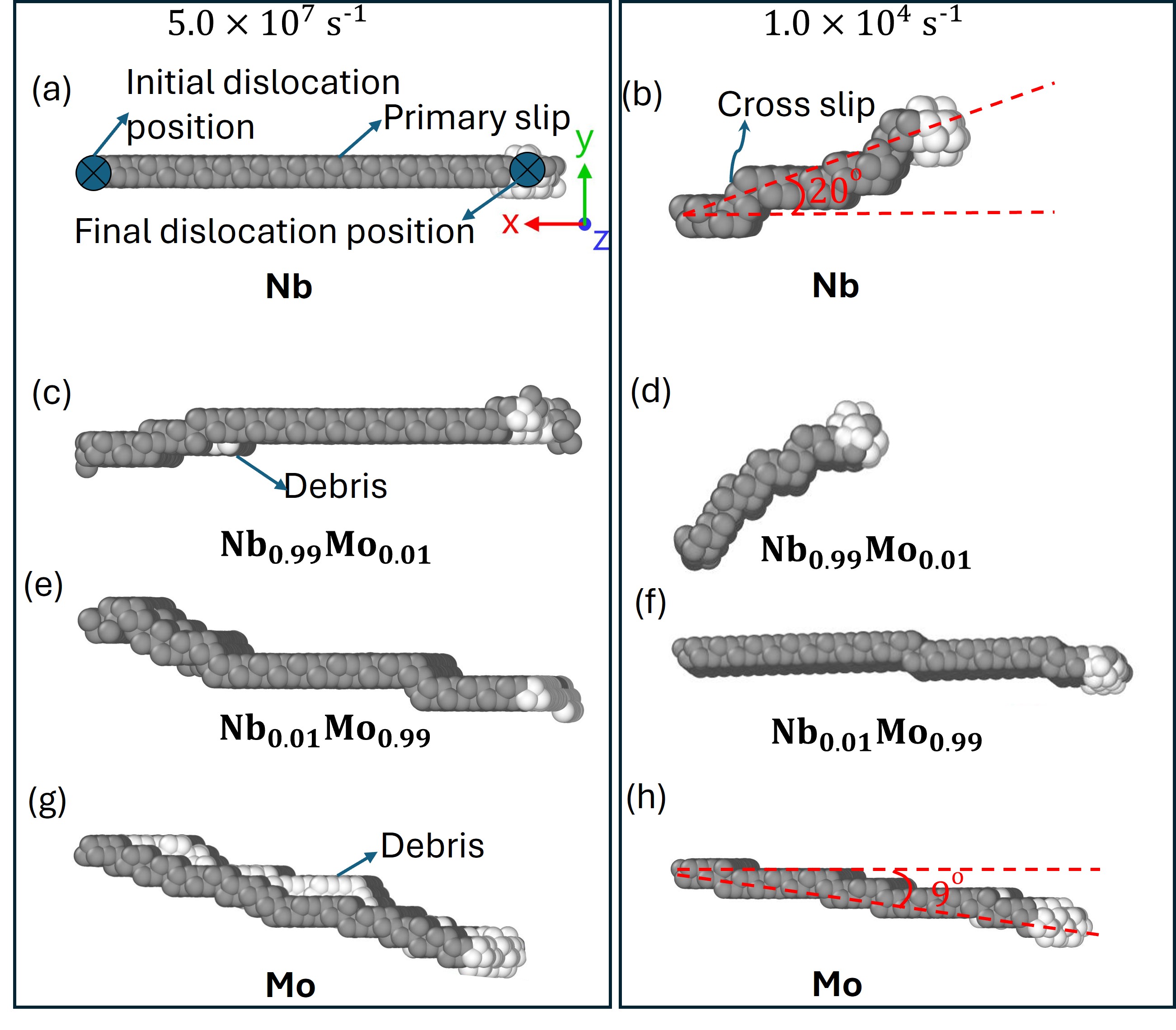}
    \caption{3D screw-dislocation slip trajectories in pure metals and their dilute alloys, constructed using the same method described in Fig.~\ref{fig:SlipPath_50nm_Nb}, viewed along the dislocation-line direction (the $z$ axis along $[\bar{1}11]$).}
    \label{fig:SlipTrajectory_50nm_PureAndDiluted} 
\end{figure}

We have performed systematic analyses of three-dimensional screw-dislocation trajectories for all cases of pure metals and dilute alloys under different strain rates, as shown in Fig.~\ref{fig:Results_50nm_Nb_and_Mo}. The results are summarized in Fig.~\ref{fig:SlipTrajectory_50nm_PureAndDiluted}. Owing to the absence of close-packed crystal planes and the nonplanar structure of screw dislocation cores in BCC metals~\cite{duesbery1998plastic,vitek2004core,Cai2004DislocationCoreEffects,dezerald2016plastic}, kink formation events can occur on distinct $\{110\}$ or $\{112\}$ slip systems. As illustrated by the mechanism schematics in Fig.~\ref{fig:GraphicalAbstract}, Fig.~\ref{fig:SlipPath_50nm_Mo_CrossKinkDragging}, and Fig.~\ref{fig:SlipPath_50nm_Mo_CrossKinkMigration}, the overall screw-dislocation trajectories are governed by the relative rates of kink nucleation and migration on both the primary slip planes (here $(\bar{1}0\bar{1})$, which experiences the maximum resolved shear stress under the applied pure shear deformation) and the cross-slip planes (other $\{110\}$ planes).

In all high–strain-rate cases (left subfigures of Fig.~\ref{fig:SlipTrajectory_50nm_PureAndDiluted}), kink migration is generally fast compared with kink nucleation on both the primary slip and cross-slip planes; thus, once a kink nucleates, it rapidly migrates along that plane. As a result, Fig.~\ref{fig:SlipTrajectory_50nm_PureAndDiluted}(a) shows nearly single-planar slip for pure Nb, reflecting very fast kink migration and relatively more difficult kink nucleation on cross-slip planes. In contrast, Fig.~\ref{fig:SlipTrajectory_50nm_PureAndDiluted}(d) shows that Mo, relative to Nb, exhibits more frequent kink nucleation on both primary and cross-slip planes. However, the greater difficulty in realigning the dislocation line following cross-slip (as described in the right subfigures of Fig.~\ref{fig:SlipPath_50nm_Mo_CrossKinkDragging}(b)) leads to enhanced cross-kink formation and the generation of larger debris via the cross-kink drag mechanism illustrated in Fig.~\ref{fig:SlipPath_50nm_Mo_CrossKinkDragging}(b). For the dilute alloys $\text{Nb}_{0.99}\text{Mo}_{0.01}$ and $\text{Nb}_{0.01}\text{Mo}_{0.99}$, the presence of solute atoms promotes additional kink nucleation on both the primary and cross-slip planes without significantly altering kink migration rates. Consequently, the trajectory of $\text{Nb}_{0.99}\text{Mo}_{0.01}$ in Fig.~\ref{fig:SlipTrajectory_50nm_PureAndDiluted}(c) shows more frequent cross-slip events and noticeable deviations from pure planar slip, accompanied by the formation of small amounts of debris due to cross-kink formation and depinning. In contrast, the trajectory of $\text{Nb}_{0.01}\text{Mo}_{0.99}$ exhibits almost no debris formation, possibly because enhanced kink nucleation on cross-slip planes facilitates rapid realignment of the dislocation line onto a common slip plane, thereby suppressing cross-kink drag events.

In all low–strain-rate cases (right subfigures of Fig.~\ref{fig:SlipTrajectory_50nm_PureAndDiluted}), kink migration is much slower, allowing relatively more frequent kink nucleation on both the primary and cross-slip planes; consequently, none of these trajectories exhibit purely planar slip features. In addition, although cross-kinks may form through the intersection of kinks on different slip planes, the lateral migration of cross-kinks revealed in Fig.~\ref{fig:SlipPath_50nm_Mo_CrossKinkMigration}(b) significantly reduces the density of point defects or debris generated during cross-kink depinning events. As a result, almost no debris is observed in any of the low–strain-rate cases. Under these kinetic conditions, the dislocation trajectory is more likely to follow an energy-minimizing path. For example, as shown in Fig.~\ref{fig:SlipTrajectory_50nm_PureAndDiluted}(b) and (h), the deviation angle between the average dislocation trajectory and the primary slip plane is approximately $20^\circ$ and $9^\circ$ for pure Nb and Mo, respectively. Both values are close to the deviation angles between the primary slip plane and the dislocation core trajectory predicted by zero-temperature density functional theory (DFT) calculations~\cite{dezerald2016plastic}. For the $\text{Nb}_{0.99}\text{Mo}_{0.01}$ alloy in Fig.~\ref{fig:SlipTrajectory_50nm_PureAndDiluted}(d), dilute Mo solute atoms promote kink nucleation on cross-slip planes, leading to more frequent cross-slip events and a further increase in the deviation angle compared with pure Nb in Fig.~\ref{fig:SlipTrajectory_50nm_PureAndDiluted}(b). In contrast, for the $\text{Nb}_{0.01}\text{Mo}_{0.99}$ alloy shown in Fig.~\ref{fig:SlipTrajectory_50nm_PureAndDiluted}(f), dilute Nb solute atoms preferentially promote kink nucleation on the primary slip plane, resulting in fewer cross-slip events and a reduced deviation angle compared with pure Mo in Fig.~\ref{fig:SlipTrajectory_50nm_PureAndDiluted}(h).

\subsection{Dislocation line length 50 nm for concentrated alloys}\label{sec:Results_50nm_b}

\begin{figure}[!htb] 
\centering
    \includegraphics[width=1.0\textwidth]{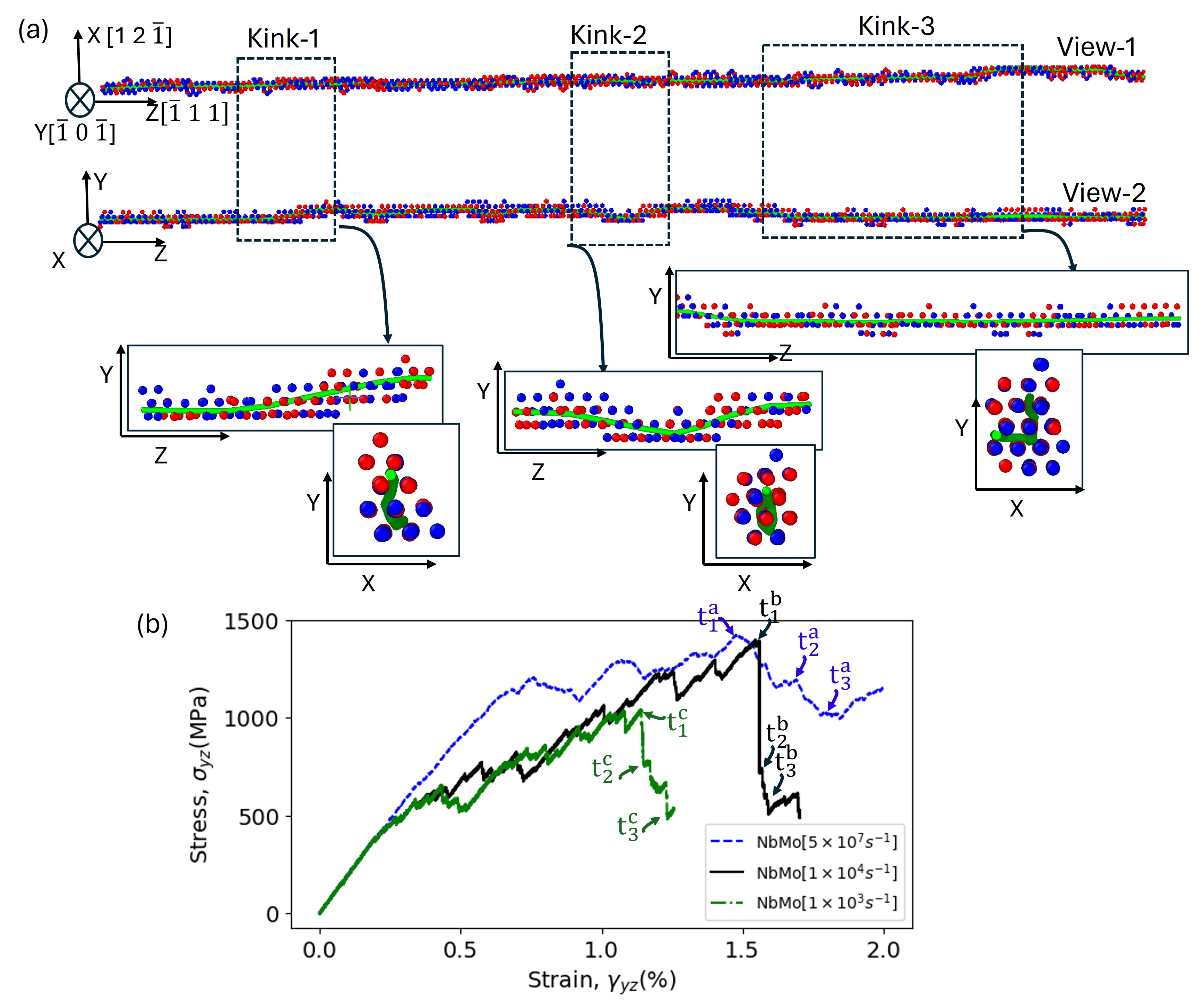}
    \caption{Simulation results for screw dislocations with a line length of 50 nm in the NbMo alloys. (a) Initial configuration of the screw dislocation after MD relaxation, shown from two viewing directions: View-1 along the $Y$ axis, projected onto the primary glide plane, and View-2 along the $X$ axis, on the plane perpendicular to the primary glide plane. Detailed views of kink and cross-kink structures at selected locations (Kink-1, Kink-2, and Kink-3) are highlighted by dashed rectangles and enlarged in the inset subfigures, which are also viewed along the $X$ and $Z$ directions. (b) Macroscopic shear stress ($\sigma_{yz}$)--strain ($\gamma_{yz}$) responses of the NbMo alloy at three different applied strain rates ($\dot{\gamma}_{yz}$) of $5\times10^{7}~\mathrm{s^{-1}}$, $1\times10^{4}~\mathrm{s^{-1}}$, and $1\times10^{3}~\mathrm{s^{-1}}$.
}
    \label{fig:StressStrain_50nm} 
\end{figure}

Figure~\ref{fig:StressStrain_50nm}(a) shows the initial stress-free configuration of the screw dislocation in the concentrated NbMo alloy. The relaxed structure is intrinsically kinked: the largest initial kinks reach a magnitude of approximately $2b$ in the direction normal to the dislocation line, as illustrated in the inset. A closer inspection reveals that these kinks are distributed on both the $(\bar{1}0\bar{1})$ and $(1\,2\,\bar{1})$ slip planes. Three major kink regions are visible (Kink-1, Kink-2, and Kink-3), highlighted by the three dashed rectangles in the inset. The left and right dashed rectangles correspond to two long single kinks connecting relatively straight dislocation segments, both lying approximately on $(1\,2\,\bar{1})$-type planes. In contrast, the central dashed rectangle shows a short double kink, where the dislocation first bends from the $(\bar{1}0\bar{1})$ plane to the $(1\,2\,\bar{1})$ plane and then bends back. As will be shown in the following MD simulations, this short double kink exhibits behavior that differs significantly from that of the two long single kinks, particularly under low-strain-rate deformation.

Figure~\ref{fig:StressStrain_50nm}(b) presents the stress–strain responses for the $50~\mathrm{nm}$ screw dislocation in the concentrated NbMo alloy under several imposed strain rates. As expected, the high–strain-rate case ($5\times10^{7}~\mathrm{s^{-1}}$) exhibits both the highest ORSS, corresponding to the initial activation of a local dislocation segment, and the highest CRSS, required to depin and move the entire dislocation line. After depinning, the stress remains high, indicating continued resistance from multiple pinning sites. In contrast, the low–strain-rate case ($1\times10^{4}~\mathrm{s^{-1}}$) shows a dramatically reduced ORSS—from above $1100~\mathrm{MPa}$ at high strain rate to below $500~\mathrm{MPa}$, a much larger reduction than observed for the 15~nm (Fig.~\ref{fig:Results15nm}) or 20~nm (Fig.~\ref{fig:Results20nm}) dislocation-line-length simulations. Although the CRSS at $1\times10^{4}~\mathrm{s^{-1}}$ is only slightly lower than that at $5\times10^{7}~\mathrm{s^{-1}}$, a key distinction is that the stress drops sharply immediately after reaching the CRSS, indicating that the entire dislocation is pinned by a single dominant pinning point. When the strain rate is further reduced to $1\times10^{3}~\mathrm{s^{-1}}$, the ORSS remains comparable to that at $1\times10^{4}~\mathrm{s^{-1}}$, but the CRSS decreases substantially, from roughly $1300~\mathrm{MPa}$ to below $1000~\mathrm{MPa}$. As in the $1\times10^{4}~\mathrm{s^{-1}}$ case, a pronounced stress drop occurs after the CRSS is reached, again indicating that a single strong pinning point governs the full-line depinning process, though with a lower CRSS at the slower rate. In the following section, we analyze the dislocation configurations immediately before and after the CRSS events to identify the nature of these strong pinning points and to clarify the strain-rate-dependent mechanisms by which the dislocation overcomes them.

\begin{figure}[!htb]
     \centering
     {\includegraphics[width=1.0\textwidth]{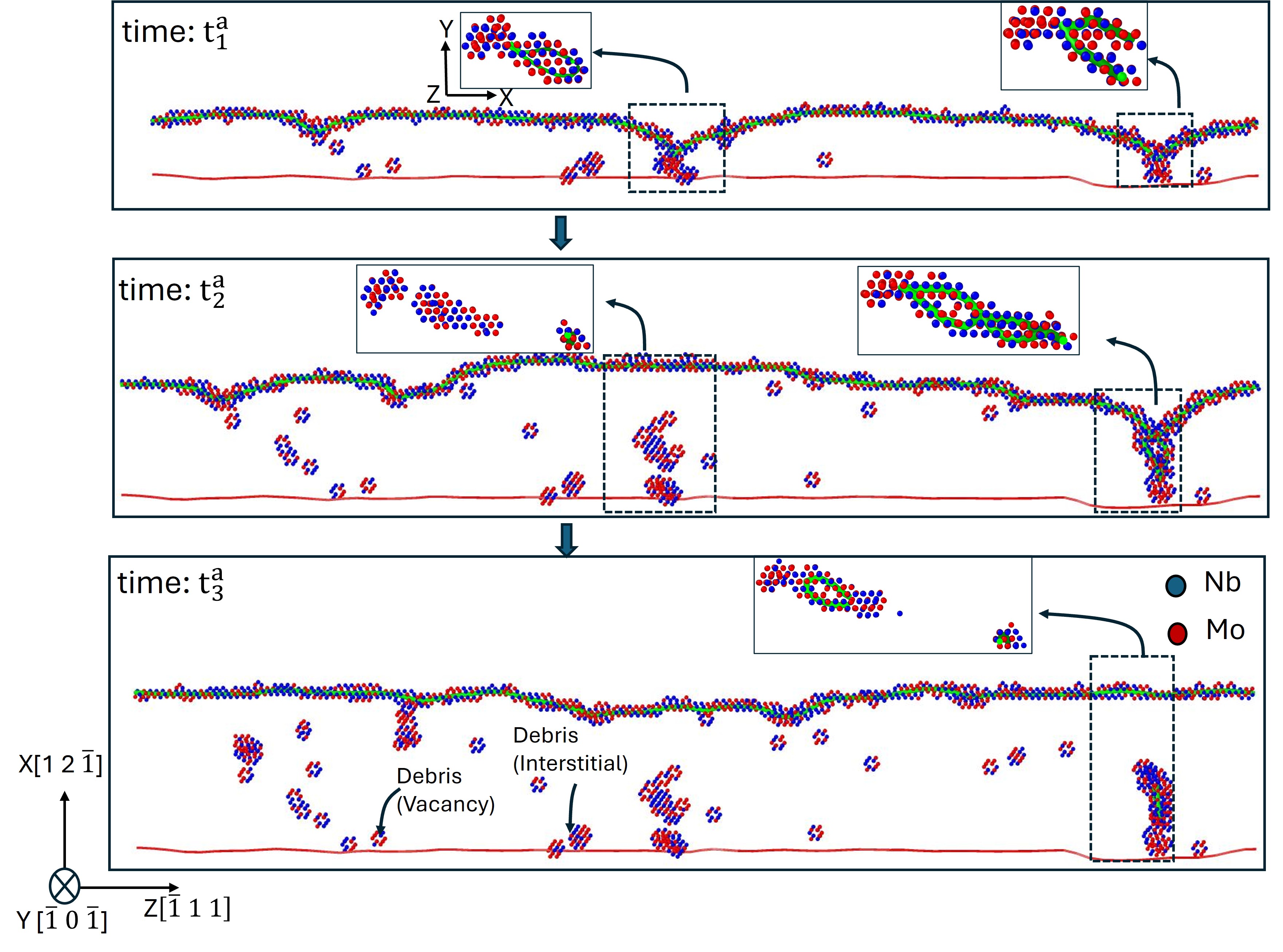}}\label{fig:ResultsCombined_50nm_5p0e7}
\caption {Snapshots of the dislocation-line configurations and associated defect structures before and after depinning for a 50~nm screw dislocation in the NbMo alloy at an applied strain rate of $\dot{\gamma}_{yz} = 5\times10^{7}~\mathrm{s^{-1}}$, corresponding to times $t_{1}^{a}$, $t_{2}^{a}$, and $t_{3}^{a}$ in Fig.~\ref{fig:StressStrain_50nm}(b). Detailed views of critical depinning events are highlighted by dashed rectangles and enlarged in the inset solid-rectangle subfigures, which are viewed along the $Z$ directions (the dislocation line direction).}
\label{fig:DislocationConfig_50nm_5p0e7}
\end{figure}

Figure~\ref{fig:DislocationConfig_50nm_5p0e7} shows the dislocation configurations immediately before and after the CRSS event at the high strain rate of $5\times10^{7}~\mathrm{s^{-1}}$. At time $t_{1}^a$, labeled on the stress–strain curve in Fig.~\ref{fig:StressStrain_50nm}(b), the dislocation line contains two strong pinning points (highlighted by dashed rectangles), both located on the original dislocation position indicated by the solid red reference line. The inset images reveal that, at each pinning site, local dislocation segments occupy different slip planes, indicating cross-kink formation. A weaker pinning point is also visible on the left side as a small curved segment formed after partial forward motion. At time $t_{2}^a$, immediately after reaching the CRSS, the applied stress drops sharply as the dislocation depins from the central pinning point (middle dashed rectangle). This depinning event produces vacancy–interstitial debris and occurs at a location that closely corresponds to the short double kink (Kink-2) in the relaxed initial configuration shown in Fig.~\ref{fig:StressStrain_50nm}(a). The debris pattern indicates that the dislocation trajectory deviates from the $(\bar{1}0\bar{1})$ plane, resembling the nonplanar glide observed in pure Mo at high strain rate.

At time $t_{3}^a$, the stress decreases further as the dislocation escapes from the remaining strong pinning point, again generating debris that is not aligned with the $(\bar{1}0\bar{1})$ plane. In this case, the debris results from a loop-emission mechanism that leaves behind a small dislocation loop on the $(\bar{1}11)$ plane, normal to the Burgers vector. Extensive vacancy and interstitial formation during glide leads to a dense debris field between the original and current dislocation positions, consistent with the high ORSS, high CRSS, and elevated post-CRSS stress in the stress–strain curve. These results closely resemble those for pure Mo shown in Fig.~\ref{fig:Results_50nm_Nb_and_Mo}(c), where cross-kink formation and drag generate large defect clusters. A direct comparison further indicates that the debris in concentrated NbMo is generally larger than in pure Mo, suggesting stronger cross-kink drag effects in the alloy. This enhanced behavior likely arises from the more complex three-dimensional energy landscape in concentrated alloys, which hinders realignment of the dislocation line onto a single primary slip plane.

%\begin{figure}[!ht] 
%\centering
%    \includegraphics[width=1.0\textwidth]{Figures/SlipPath_50nm_NbMo_5p01e7.jpg}
%    \caption{Dislocation slip path of NbMo at time $t^a_3$ for $\dot{\gamma}_{yz} = %5.0\times10^{7}\ \mathrm{s^{-1}}$}
%    \label{fig:SlipPath_50nm_NbMo_5p01e7} 
%\end{figure}

%We plot the three-dimensional screw-dislocation trajectories based on the deformation gradient $F_{zy}$ for NbMo at time $t^{a}_{3}$ under the high–strain-rate condition in Fig.~\ref{fig:Results_50nm_Nb_and_Mo} . These results are very similar to those shown in Fig.~\ref{fig:SlipPath_50nm_Mo_CrossKinkDragging}(a) for pure Mo with the same dislocation line length and strain rate, where large clusters of interstitials and vacancies are generated due to the formation and drag of cross-kinks, as described in Fig.~\ref{fig:SlipPath_50nm_Mo_CrossKinkDragging}(b). A direct comparison further shows that the debris formed in the concentrated NbMo alloy is generally larger than that in pure Mo, indicating stronger cross-kink drag effects in the concentrated alloy. This behavior likely arises because screw dislocation glide in concentrated alloys occurs within a more complex three-dimensional energy landscape, making realignment of the entire dislocation line onto a single primary slip plane more difficult, as illustrated in the right subfigures of Fig.~\ref{fig:SlipPath_50nm_Mo_CrossKinkDragging}(b). As a result, cross-kink drag events occur more frequently, leading to the formation of larger debris clusters.

\begin{figure}[!htb]
     \centering
     {\includegraphics[width=1.0\textwidth]{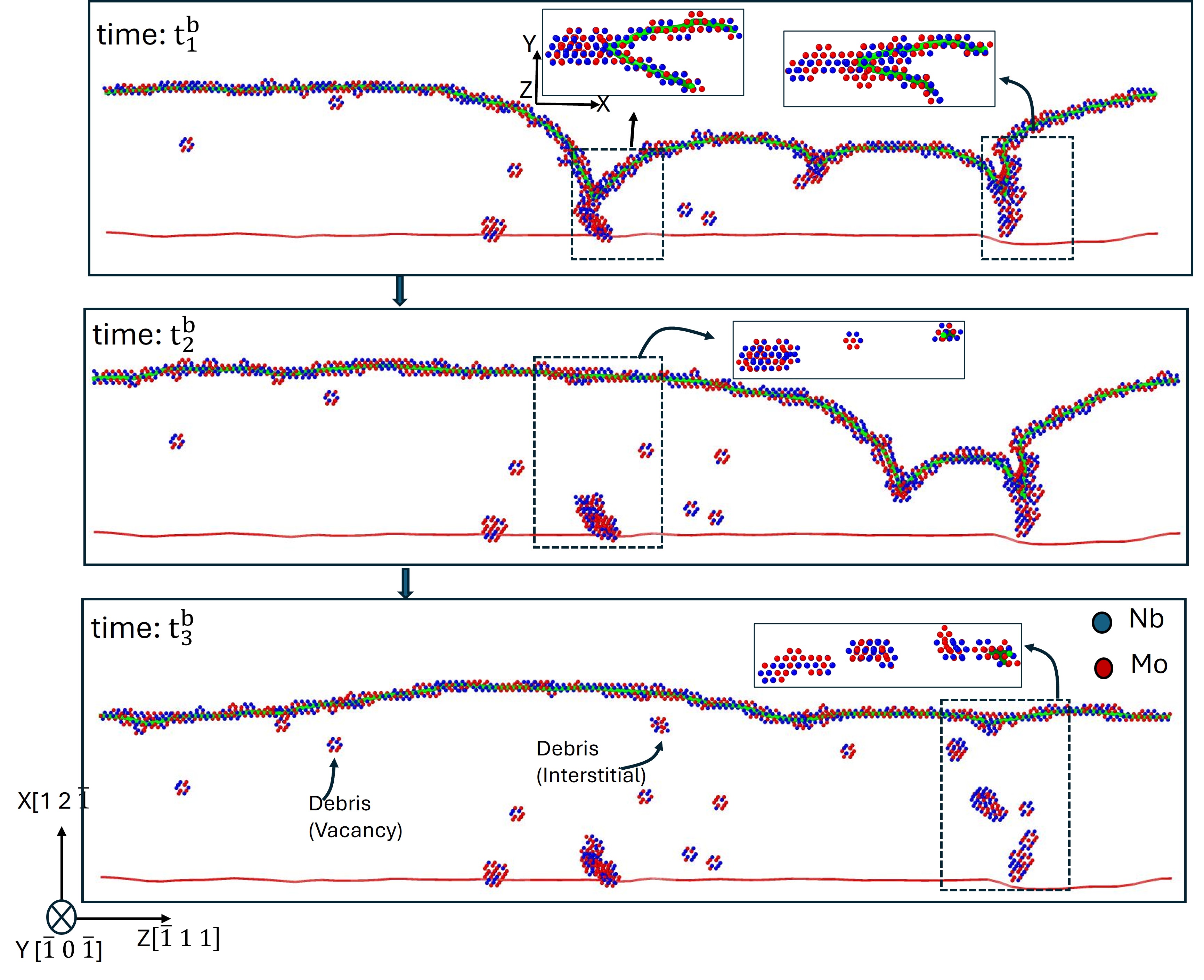}}\label{fig:ResultsCombined_50nm_1p0e4}
\caption {Snapshots of the dislocation-line configurations and associated defect structures before and after depinning for a 50~nm screw dislocation in the NbMo alloy at an applied strain rate of $\dot{\gamma}_{yz} = 1\times10^{4}~\mathrm{s^{-1}}$, corresponding to times $t_{1}^{b}$, $t_{2}^{b}$, and $t_{3}^{b}$ in Fig.~\ref{fig:StressStrain_50nm}(b). Detailed views of critical depinning events are highlighted by dashed rectangles and enlarged in the inset solid-rectangle subfigures, which are viewed along the $Z$ directions (the dislocation line direction).}
\label{fig:DislocationConfig_50nm_1p0e4}
\end{figure}

Figure~\ref{fig:DislocationConfig_50nm_1p0e4} shows the dislocation configurations immediately before and after the CRSS event for the lower strain rate of $1\times10^{4}~\mathrm{s^{-1}}$, corresponding to times $t_{1}^b$, $t_{2}^b$, and $t_{3}^b$ in Fig.~\ref{fig:StressStrain_50nm}(b). Similar to the high–strain-rate case (Fig.~\ref{fig:DislocationConfig_50nm_5p0e7}), two strong pinning points are observed at the same locations along the original dislocation line (marked by the solid red reference line) and highlighted by dashed rectangles at time $t_{1}^b$. The inset images confirm that local dislocation segments at these sites lie on different slip planes, indicating cross-kink formation. However, the local curvature around these pinning regions is significantly larger than in the high–strain-rate case, suggesting that adjacent segments have moved farther from their original positions. A weaker pinning point is also present between the two strong ones, resulting from local glide. At time $t_{2}^b$, immediately after reaching the CRSS, the stress drops sharply as the dislocation depins from the central pinning point, which again corresponds to the short double kink (Kink-2) in Fig.~\ref{fig:StressStrain_50nm}(a). Although vacancy–interstitial debris is generated, it is much smaller in volume than in the high–strain-rate case and remains predominantly aligned with the $(\bar{1}0\bar{1})$ plane. At time $t_{3}^b$, further stress reduction occurs as the dislocation escapes from the remaining pinning point, producing additional debris that also remains nearly planar. Overall, both the number of cross-kinks and the amount of debris are substantially reduced compared with the high–strain-rate case, consistent with the lower post-depinning stress levels. These results indicate that at lower strain rates the system has sufficient time to reorganize local dislocation segments and relax cross-kink structures, allowing the dislocation to overcome strong pinning points under significantly lower applied stress.

\begin{figure}[!htb] 
\centering
    \includegraphics[width=1.0\textwidth]{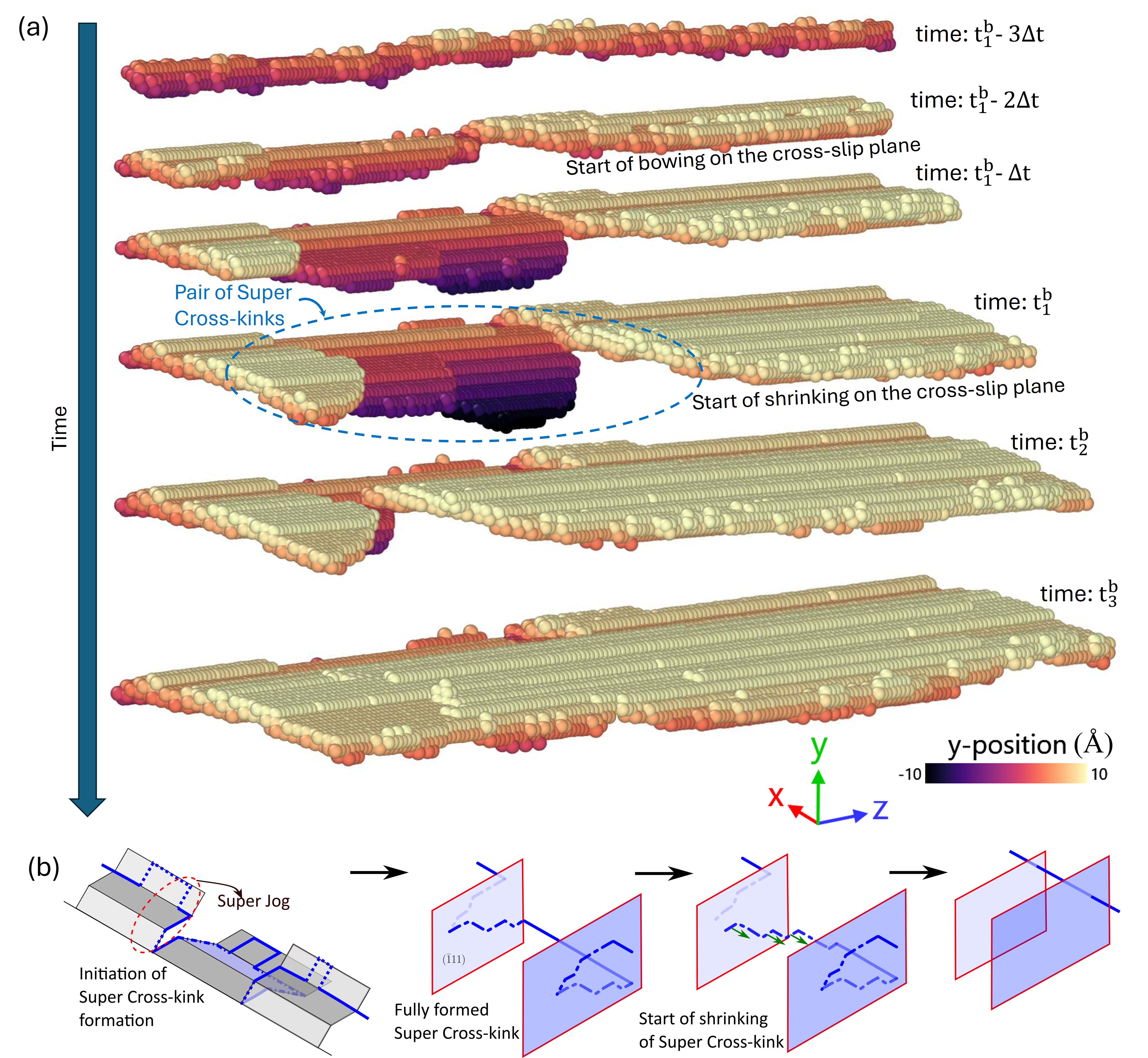}
    \caption{An alternative depinning mechanism from cross-kinks observed in the NbMo alloy at a strain rate of $\dot{\gamma}_{yz} = 1.0\times10^{4}\ \mathrm{s^{-1}}$. (a) Snapshots of the three-dimensional screw-dislocation slip trajectories, plotted using the same method described in Fig.~\ref{fig:SlipPath_50nm_Nb}, corresponding to times $t_{1}^{b}$, $t_{2}^{b}$, and $t_{3}^{b}$ in Fig.~\ref{fig:StressStrain_50nm}(b). In this visualization, the selected atoms are colored according to their relative coordinates along the $y$ axis (normal to the $(\bar{1}0\bar{1})$ plane), with the initial dislocation-line position taken as the reference zero. Dynamic evolution of the slip path can be found in the Supplementary Materials \textbf{Video S1}. (b) Schematic illustration of the atomistic configurations shown in (a), based on dislocation-line segment representations, highlighting the formation of a pair of super-cross-kinks and their subsequent depinning via lateral kink migration (a 3D forward–backward cross-slip process).}
    \label{fig:SlipPath_50nm_NbMo_1p0e4} 
\end{figure}

For this low–strain-rate case, we also plot the three-dimensional screw-dislocation trajectories for NbMo based on the deformation gradient $F_{zy}$ in Fig.~\ref{fig:SlipPath_50nm_NbMo_1p0e4}(a). The selected atoms are colored according to their relative coordinates along the $y$ axis, following the same convention as in Fig.~\ref{fig:SlipPath_50nm_Mo_CrossKinkMigration}(b). These three-dimensional trajectories provide a clearer view of cross-kink formation and depinning than the two-dimensional dislocation-line configurations shown in Fig.~\ref{fig:DislocationConfig_50nm_1p0e4}. Before time $t_{1}^{b}$, corresponding to the CRSS in Fig.~\ref{fig:StressStrain_50nm}(b), a large segment of the dislocation line bows out onto a cross-slip plane, as indicated by the darker-colored atoms at times $t_{1}^{b}-2\Delta t$ and $t_{1}^{b}-\Delta t$. At time $t_{1}^{b}$, two major segments are present: one gliding on the primary slip plane (light yellow atoms) and the other on the cross-slip plane (dark red, purple, and black atoms). Their intersection gives rise to a pair of large cross-kink structures near the junction points, referred to here as \emph{super cross kinks} (dashed circle at $t_{1}^{b}$). This moment coincides with the attainment of the CRSS and the onset of retraction of the cross-slip segment back toward the primary slip plane, accompanied by a rapid stress drop. At time $t_{2}^{b}$, most of the cross-slip segment has retracted and begins to glide back onto the primary plane. By time $t_{3}^{b}$, the cross-slip segment has fully disappeared, and the entire dislocation glides on the primary slip plane, leaving only a small amount of debris.

To further clarify this mechanism, particularly the formation and elimination of the super cross kinks, we present a schematic illustration in Fig.~\ref{fig:SlipPath_50nm_NbMo_1p0e4}(b). The leftmost subfigure shows the initiation of super cross-kink formation caused by dislocation segments cross-slipping onto different slip planes. The middle-left subfigure depicts a pair of fully developed super cross kinks formed on two parallel $(\bar{1}11)$ planes, highlighted by the red parallelograms. In the middle-right subfigure, multiple kinks on cross-slip planes migrate along the dislocation-line direction ($Z \parallel [\bar{1}11]$), as indicated by the green arrows. When these kink migrations traverse the entire cross-slip segment, the super cross kink is eliminated and the dislocation line returns to the original slip plane, as shown in the rightmost subfigure.

\begin{figure}[!htb]
     \centering
     {\includegraphics[width=1.0\textwidth]{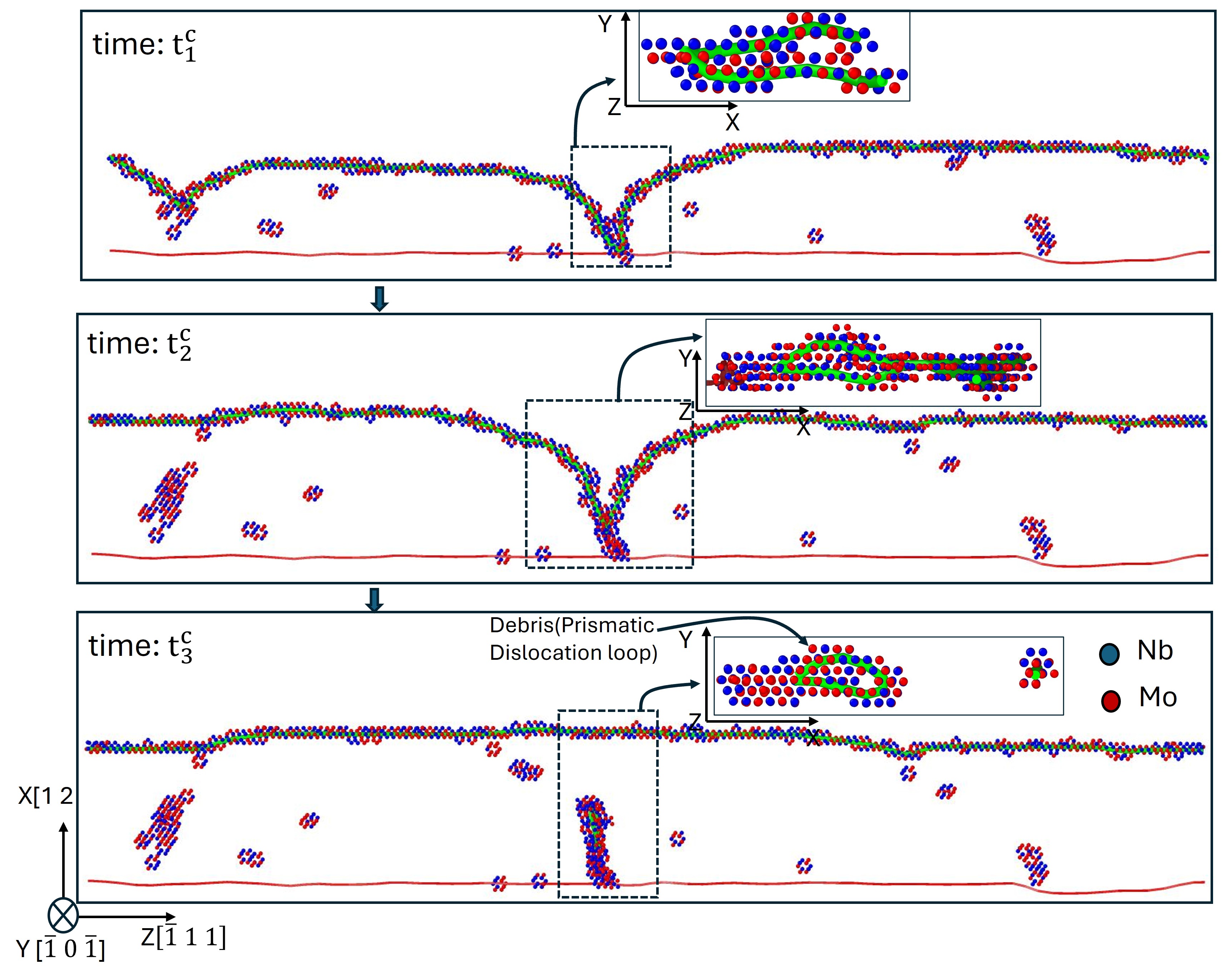}}\label{fig:ResultsCombined_50nm_1p0e3}
\caption {Snapshots of the dislocation-line configurations and associated defect structures before and after depinning for a 50~nm screw dislocation in the NbMo alloy at an applied strain rate of $\dot{\gamma}_{yz} = 1\times10^{3}~\mathrm{s^{-1}}$, corresponding to times $t_{1}^{c}$, $t_{2}^{c}$, and $t_{3}^{c}$ in Fig.~\ref{fig:StressStrain_50nm}(b). Detailed views of critical depinning events are highlighted by dashed rectangles and enlarged in the inset solid-rectangle subfigures, which are viewed along the $Z$ directions (the dislocation line direction).}
\label{fig:DislocationConfig_50nm_1p0e3}
\end{figure}

Figure~\ref{fig:DislocationConfig_50nm_1p0e3} shows the dislocation configurations immediately before and after the CRSS event for the lowest strain rate of $1\times10^{3}~\mathrm{s^{-1}}$, corresponding to times $t_{1}^c$, $t_{2}^c$, and $t_{3}^c$ in Fig.~\ref{fig:StressStrain_50nm}(b). In contrast to the higher strain-rate cases shown in Figs.~\ref{fig:DislocationConfig_50nm_5p0e7} and \ref{fig:DislocationConfig_50nm_1p0e4}, only a single strong pinning point is present when the stress reaches the CRSS at time $t_{1}^c$. This pinning site, highlighted by the dashed rectangle, is located near the short double kink (Kink-2) identified in Fig.~\ref{fig:StressStrain_50nm}(a). A weaker pinning point also appears on the left side of the dislocation line after partial glide away from the original position. At time $t_{2}^c$, the stress decreases slightly as the dislocation first depins from the weaker obstacle, while the central strong pinning point remains at the original dislocation position (marked by the solid red line). Between $t_{2}^c$ and $t_{3}^c$, this strong pinning point is eventually overcome through a loop-emission mechanism, leaving behind a small dislocation loop on the $(\bar{1}11)$ plane normal to the Burgers vector. These results indicate that at very low strain rates the system has sufficient time to reorganize local dislocation segments and relax cross-kink structures, thereby reducing the number of strong pinning sites. The increased spacing between pinning points makes local bending and loop emission more favorable than direct cutting, resulting in minimal vacancy–interstitial debris and a lower applied stress required for sustained dislocation glide.

\begin{figure}[!htb] 
\centering
    \includegraphics[width=1.0\textwidth]{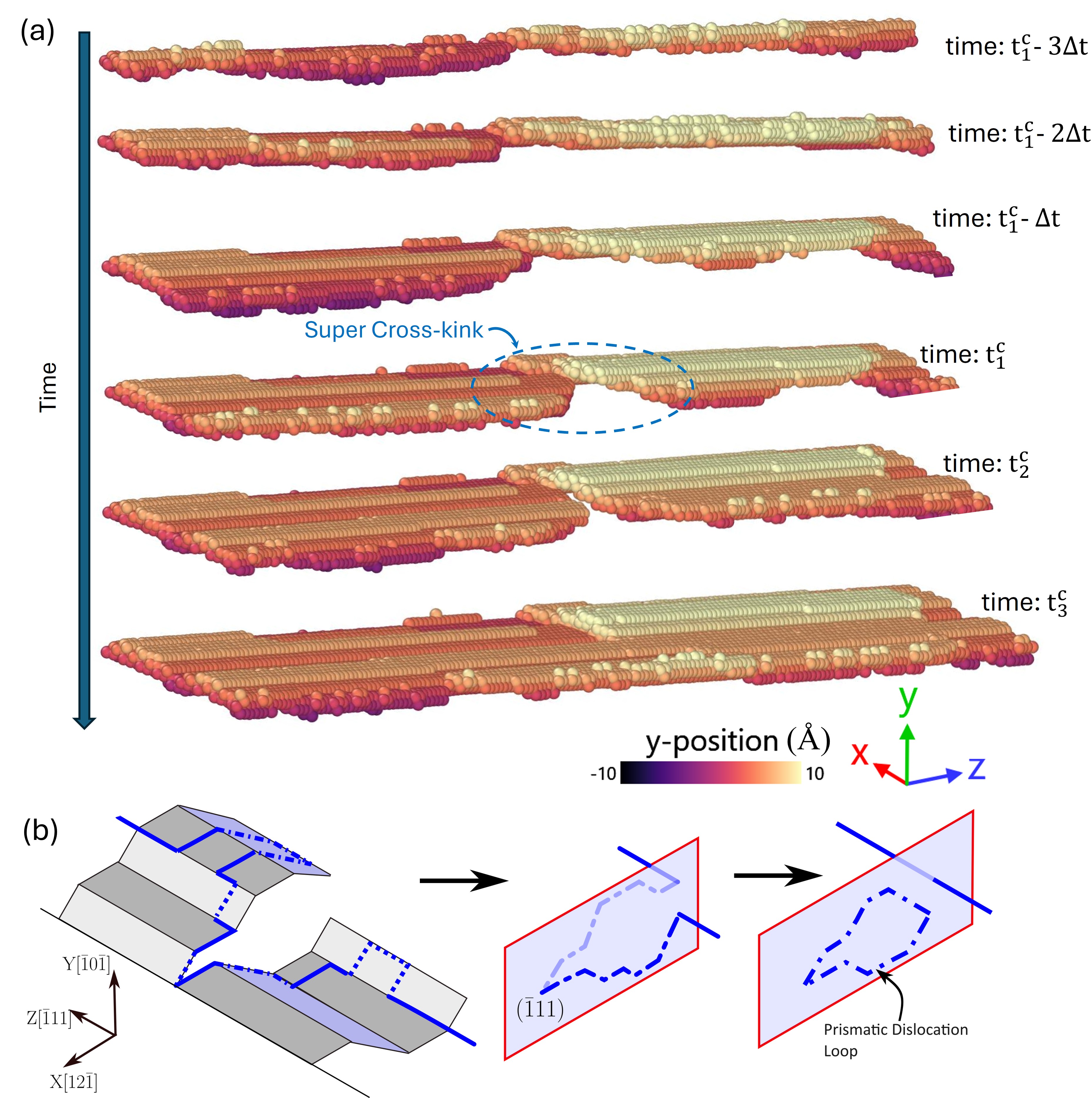}
    \caption{An alternative depinning mechanism from cross-kinks observed in the NbMo alloy at a strain rate of $\dot{\gamma}_{yz} = 1.0\times10^{3}\ \mathrm{s^{-1}}$. (a) Snapshots of the three-dimensional screw-dislocation slip trajectories, plotted using the same method described in Fig.~\ref{fig:SlipPath_50nm_Nb}, corresponding to times $t_{1}^{c}$, $t_{2}^{c}$, and $t_{3}^{c}$ in Fig.~\ref{fig:StressStrain_50nm}(b). The color is based on the same method of Fig. \ref{fig:SlipPath_50nm_NbMo_1p0e4}(a). Dynamic evolution of the slip path can be found in the Supplementary materials \textbf{Video S2}. (b) Schematic illustration of the atomistic configurations shown in (a), based on dislocation-line segment representations, highlighting the formation of a single super cross-kink and the subsequent depinning via a prismatic loop formation process~\cite{swinburne2016fast}.}
    \label{fig:SlipPath_50nm_NbMo_1p0e3} 
\end{figure}

For this lowest–strain-rate case, we also plot the three-dimensional screw-dislocation trajectories for NbMo based on the deformation gradient $F_{zy}$ in Fig.~\ref{fig:SlipPath_50nm_NbMo_1p0e3}(a). The atoms are colored according to their relative coordinates along the $y$ axis, following the same convention as in Fig.~\ref{fig:SlipPath_50nm_Mo_CrossKinkMigration}(b) and Fig.~\ref{fig:SlipPath_50nm_NbMo_1p0e4} (a). From time $t_{1}^{c}-3\Delta t$ to $t_{1}^{c}-\Delta t$, a segment of the dislocation line bows out onto a cross-slip plane, as indicated by atoms colored orange to dark red. However, unlike the behavior in Fig.~\ref{fig:SlipPath_50nm_NbMo_1p0e4}, the bowing remains limited. At time $t_{1}^{c}$—corresponding to the CRSS in Fig.~\ref{fig:StressStrain_50nm}(b)—two major dislocation segments are already present: one on a primary slip plane (light yellow atoms) and the other on a neighboring primary slip plane (orange and light red atoms). Their junction forms a single super cross kink (highlighted by the dashed circle), in contrast to the pair of super cross kinks observed at higher strain rates in Fig.~\ref{fig:SlipPath_50nm_NbMo_1p0e4}. Between $t_{1}^{c}$ and $t_{2}^{c}$, the segment on the primary slip plane begins to cross-slip onto the adjacent plane. By time $t_{3}^{c}$, most of the dislocation line has returned to a single primary slip plane, leaving behind a small dislocation loop, as shown in Fig.~\ref{fig:DislocationConfig_50nm_1p0e3}. These results indicate that at sufficiently low strain rates, although cross-slip events occur on multiple crystallographic planes, the elastic energy of the dislocation line suppresses large bowing across different planes and favors a predominantly planar configuration.

A schematic illustration of this mechanism is presented in Fig.~\ref{fig:DislocationConfig_50nm_1p0e3}(b). The leftmost subfigure shows the formation of a single super cross kink resulting from dislocation segments lying on different slip planes. A notable feature is the onset of cross-slip back to the original primary slip plane, initiated by double-kink nucleation at the leading edge of these segments. Once formed, the kinks migrate along the dislocation-line direction and consolidate into a super cross kink on a $(\bar{1}11)$ plane, as depicted in the middle subfigure, where the incipient loop is also visible. Continued cross-slip onto the primary slip plane eventually closes the loop on the $(\bar{1}11)$ plane, forming a prismatic dislocation loop (right subfigure)~\cite{swinburne2016fast}. The long dislocation line subsequently depins from the super cross kink and resumes glide.

\section{Discussion}
\label{sec:Discussion}

\subsection{Rate-dependent depinning mechanisms: beyond defect-assisted cutting}

The present simulations demonstrate that cross-kink depinning in BCC metals and alloys is strongly strain-rate- and line-length-dependent, and that defect-assisted cutting is not the only operative mechanism. At high strain rates ($\sim 10^{7}\ \mathrm{s^{-1}}$), in all investigated cases with different dislocation lengths in the simulation supercells, depinning is frequently accompanied by the formation of vacancy--interstitial clusters, even in pure Mo. In this regime, multiple cross-kinks accumulate before the dislocation can reorganize elastically, and their removal proceeds through drag or cutting processes that leave permanent debris, as illustrated in Fig. \ref{fig:GraphicalAbstract}(b) and detailed in Fig. \ref{fig:SlipPath_50nm_Mo_CrossKinkDragging}(b). This behavior is broadly consistent with Suzuki-type and subsequent models, where jog or cross-kink depinning through the cutting/drag process is intrinsically associated with point-defect generation~\cite{suzuki1980solid,Hattendorf92,Rao2019ModelingSolutionHardening,Maresca2020TheoryScrewDislocation,Rao2021TheorySolidSolution,Ghafarollahi2022ScrewcontrolledStrengthBCC}.

In contrast, at lower strain rates ($10^{3}$--$10^{4}\ \mathrm{s^{-1}}$), alternative depinning pathways become accessible. Cross-kinks can migrate laterally along the dislocation line (Fig. \ref{fig:SlipPath_50nm_Mo_CrossKinkMigration}(b)), shrink via elastic relaxation, or be eliminated through three-dimensional forward--backward glide (Fig. \ref{fig:SlipPath_50nm_NbMo_1p0e4}(b)) without producing significant point defects. These mechanisms are increasingly prominent for longer dislocation lines, where enhanced elastic line-tension effects and configurational freedom allow collective rearrangements prior to catastrophic failure. Thus, defect-assisted cutting represents only one branch of a broader depinning landscape. Whether cross-kinks fail by cutting or by point-defect-free reconfiguration is dynamically selected by the competition among kink nucleation, kink migration, elastic line tension, and the time available for thermally activated relaxation under the imposed strain rate.

\subsection{Cross-kink formation and characteristic obstacle spacing}

In concentrated NbMo alloys, the relaxed screw dislocation configuration already exhibits intrinsic kinked morphology, consistent with the equilibrium roughening picture proposed in statistical-mechanical models~\cite{Maresca2020TheoryScrewDislocation}. However, our simulations reveal that the initial equilibrium kinks are not necessarily the strength-determining obstacles. Under applied shear, the dislocation configuration evolves further, and local segments can develop into larger multi-plane jog structures. Collisions between such extended jogs give rise to what may be termed \emph{super-cross-kinks}, i.e., cross-kink structures that span multiple atomic layers across adjacent $\{110\}$ planes and exceed the depth of ordinary cross-kinks, as shown in Figs. \ref{fig:SlipPath_50nm_NbMo_1p0e4} and \ref{fig:SlipPath_50nm_NbMo_1p0e3}. These super-cross-kinks behave as particularly strong pinning points. Their depinning consistently coincides with the macroscopic CRSS in long-line simulations and is frequently associated with either (i) the emission of prismatic dislocation loops or (ii) collective lateral migration of the super-cross-kink structure. While existing Suzuki-, Rao-, and Curtin-type frameworks correctly identify cross-kinks or jogs as dominant obstacles and commonly invoke defect-assisted cutting, the present results indicate that, in concentrated alloys, strengthening may be controlled by the failure of such extended super-cross-kinks rather than by isolated elementary jogs.

The emergence of these obstacles also refines the concept of characteristic obstacle spacing, which controls the stress required for dislocation bowing. In Suzuki’s classical theory, the mean jog spacing is selected dynamically through minimization of the total flow stress~\cite{Suzuki1979SolidSolutionHardening,suzuki1980solid,Hattendorf92}. Rao’s extension incorporates explicit solute–core interactions within a stress-based jog-dragging formulation~\cite{Rao2019ModelingSolutionHardening,Rao2021TheorySolidSolution}. In the statistical-mechanical framework of Maresca and Curtin, equilibrium roughening yields a characteristic cross-kink spacing that scales with the intrinsic kink length $\zeta_c$~\cite{Maresca2020TheoryScrewDislocation}.  Our atomistic results indicate that the effective obstacle spacing in refractory concentrated alloys is neither purely stress-selected nor purely equilibrium-selected, as discussed below. 

Equilibrium roughening establishes a population of potential pinning sites in the relaxed configuration, as observed in Fig.~\ref{fig:StressStrain_50nm}(a). In particular, the central Kink-2 consistently acts as a strong pinning point under all examined strain-rate conditions. During deformation, this site repeatedly evolves into a super-cross-kink that spans multiple adjacent slip planes and governs the macroscopic CRSS, independent of strain rate. Its presence can already be inferred from thermodynamically roughened configurations prior to loading, suggesting that certain hard pinning points are embedded in the equilibrium dislocation morphology. However, applied stress and strain rate further modulate the evolution of these sites. At low strain rates, many transient cross-kinks are eliminated through elastic reconfiguration, leaving only persistent hard obstacles such as Kink-2. At high strain rates, limited relaxation promotes the accumulation and defect-assisted cutting of cross-kinks, effectively reducing the kinetic obstacle spacing. Thus, the strength-determining spacing emerges from coupled thermodynamic roughening and kinetic amplification processes. At present, a rigorous criterion for identifying such persistent hard pinning points from equilibrium configurations alone remains lacking. Developing quantitative descriptors to distinguish transient kinks from strength-dominant pinning points in the presence of local chemical fluctuations is an important direction for future theoretical and atomistic studies.

\subsection{Chemical effects and their influence on kink behavior}

The glide behavior in pure Nb, pure Mo, and their dilute alloys (Fig.~\ref{fig:SlipTrajectory_50nm_PureAndDiluted}) reveals important chemistry-dependent trends. In pure Nb, Mo, and their dilute counterparts, the relaxed screw dislocation configuration is initially straight, without intrinsic kinks. However, their subsequent glide mechanisms differ markedly. Pure Nb and Nb-rich dilute alloys exhibit strong non-Schmid effects, leading to frequent cross-slip onto other $\{110\}$-type planes. This behavior is consistent with the dislocation core topology in Nb, where the transition pathway between adjacent easy-core configurations involves a pronounced deviation of the intermediate hard-core configuration~\cite{Dezerald2016PlasticAnisotropyDislocation}. As a result, cross-slip is more readily activated in Nb in low strain rate cases.

In contrast, pure Mo shows comparatively fewer cross-slip events but a higher tendency to form cross-kinks once multi-plane glide segments are established. At high strain rates, Mo exhibits a stronger propensity for debris formation (vacancies and self-interstitial clusters), and such defect generation is consistently preceded by cross-kink formation. This suggests that Mo, although less prone to frequent cross-slip than Nb, more readily stabilizes cross-kink geometries that evolve into defect-producing depinning events. The difference can be traced to the relative positioning of easy and hard core configurations in Mo, where the intermediate hard-core state lies nearly along the straight line connecting adjacent easy cores~\cite{Dezerald2016PlasticAnisotropyDislocation}, modifying the energetic pathway for kink evolution.

The addition of small amounts of Mo into Nb, or Nb into Mo, introduces local chemical heterogeneity and lattice distortion that facilitate double-kink nucleation. In dilute regimes, this generally increases kink activity while reducing the probability that cross-kinks evolve into large debris-producing structures. However, at equiatomic concentration, the chemically rough energy landscape becomes sufficiently complex that even the relaxed configuration is kinked, and subsequent evolution under stress generates highly complex 3D structures, including super-cross-kinks spanning multiple glide planes.

\subsection{Implications for strengthening models and future directions}

These results suggest several refinements to existing strengthening theories. First, cross-kinks should be treated as a heterogeneous population, including ordinary jogs and extended super-cross-kinks that span multiple slip planes. Second, the effective obstacle spacing entering CRSS scaling relations should not be viewed as a fixed quantity, but rather as strain-rate- and line-length-dependent, emerging from the interplay between equilibrium roughening and subsequent kinetic evolution under load. Third, chemical effects enter not only through average solute strength but also through their influence on dislocation core structures, cross-slip probability, and multi-plane kink interactions.

The present large-scale kinetic simulations employ the ADP potential of Starikov \emph{et al.}~\cite{Starikov2024AngulardependentInteratomicPotential}, which is well suited for capturing screw-dislocation energetics and mobility trends at moderate computational cost. Nevertheless, more accurate machine-learning interatomic potentials~\cite{zheng2023multi,mamun2023comparing,wang2024unraveling,li2025dislocation}, carefully benchmarked against first-principles calculations of dislocation core structures, Peierls barriers/stress on multiple types of slip planes, and non-Schmid effects~\cite{duesbery1998plastic,vitek2004core,Dezerald2016PlasticAnisotropyDislocation,kraych2019non,romero2022atomistic}, would enable more precise molecular statics analyses of equilibrium kink populations and more rigorous statistical characterization of strength-determining obstacle spacing at low strain rates. Such developments will be particularly important for identifying equilibrium-selected strong pinning points under realistic chemical short-range order (SRO), medium-range order (MRO), or nanoscale heterogeneity, which should themselves be generated using accurate Monte Carlo and mesoscale simulations~\cite{han2024ubiquitous,xi2024kinetic,islam2025nonequilibrium,xi2025multiscale}.

Overall, while Suzuki-, Rao-, and Curtin-type frameworks correctly identify cross-kinks or jogs as dominant strengthening obstacles and commonly treat depinning through defect-assisted processes~\cite{suzuki1980solid,Hattendorf92,Rao2019ModelingSolutionHardening,Maresca2020TheoryScrewDislocation,Rao2021TheorySolidSolution,Ghafarollahi2022ScrewcontrolledStrengthBCC}, the present atomistic evidence demonstrates that (i) multiple depinning pathways coexist, (ii) certain strong pinning points arising from the chemically complex energy landscape can evolve into extended cross-kink structures that govern the CRSS, and their depinning can proceed through complex 3D dislocation-line rearrangements, including those described in Fig.~\ref{fig:SlipPath_50nm_NbMo_1p0e4}(b) and Fig.~\ref{fig:SlipPath_50nm_NbMo_1p0e3}(b), and (iii) the effective spacing between strength-dominant pinning points is governed by coupled thermodynamic and kinetic effects. Incorporating these features into future strengthening models will be essential for the predictive design of refractory concentrated alloys.

\section{Conclusions}
\label{sec:Conclusion}

In this work, conventional molecular dynamics and strain-boost hyperdynamics simulations were employed to systematically investigate screw-dislocation glide, cross-kink formation, and depinning mechanisms, as illustrated in Fig.~\ref{fig:GraphicalAbstract}, in pure Nb and Mo, dilute Nb–Mo alloys, and equiatomic NbMo as a representative refractory concentrated alloy. By systematically varying both the strain rate ($10^{3}$--$10^{7}\ \mathrm{s^{-1}}$) and the dislocation line length (15--50~nm) at room temperature (300~K), we first demonstrate that low-strain-rate simulations require sufficiently long dislocation lines to capture consistent cross-kink behavior. Using the results obtained for the longest dislocation length (50~nm), we then draw the following main conclusions:

\begin{enumerate}
    \item Cross-kinks (jogs) can form not only in chemically complex alloys but also in pure BCC metals such as Mo. Their formation and evolution are governed by the relative rates of kink nucleation and migration on primary and cross-slip planes, which depend sensitively on chemical species through differences in dislocation core structure and non-Schmid behavior (e.g., the distinct glide characteristics of Nb and Mo).

    \item At high strain rates, depinning of cross-kinks through the cutting/drag process frequently proceeds through defect-assisted mechanisms involving vacancy–interstitial cluster formation (Fig.~\ref{fig:SlipPath_50nm_Mo_CrossKinkDragging}(b)), consistent with classical jog-dragging concepts.

    \item At low strain rates, additional elastic and collective depinning pathways without significant point defect generations become accessible, including lateral cross-kink migration (Fig. \ref{fig:SlipPath_50nm_Mo_CrossKinkMigration}(b)), 3D forward–backward cross-slip processes that restore planar glide (Fig.~\ref{fig:SlipPath_50nm_NbMo_1p0e4}(b)), and prismatic loop formation (Fig.~\ref{fig:SlipPath_50nm_NbMo_1p0e3}(b)). Thus, defect-assisted cutting represents only one branch of a broader depinning landscape.

    \item In concentrated NbMo alloys, specific strong pinning points embedded in the MD relaxed roughened dislocation morphology (e.g., the central Kink-2 in Fig.~\ref{fig:StressStrain_50nm}(a)) consistently control the CRSS. Under load, such sites may evolve into extended multi-plane cross-kink structures (\emph{super cross-kinks}); however, these extended configurations are consequences of the underlying hard pinning points rather than their fundamental origin.

    \item The effective obstacle spacing governing dislocation bowing strength is therefore neither purely equilibrium-selected nor purely stress-selected. At low strain rates, it is more closely associated with certain persistent strong pinning points that survive after MD relaxations, whereas at high strain rates, it is influenced by kinetically amplified cross-kink accumulation. Developing quantitative criteria to define and identify such strength-dominant pinning points from the relaxed dislocation configurations remains an important direction for future work.
\end{enumerate}

Overall, these findings demonstrate that cross-kink-controlled strengthening in BCC metals and refractory concentrated alloys is intrinsically chemistry-, rate-, and length-scale dependent. The operative depinning mechanisms and the strength-determining obstacle spacing emerge from a coupled competition among kink nucleation, kink migration, non-Schmid effects, elastic line tension, and thermally activated relaxation. Incorporating these coupled effects into physically based strengthening models will be essential for improving predictive capability in refractory concentrated alloys and related BCC systems.

\section*{Acknowledgments}
S.C. and L.Q. gratefully thank the funding support from National Science Foundation, United States, grant \#DMR-1847837. The calculations were performed by using the Extreme Science and Engineering Discovery Environment (XSEDE) Stampede3 at the TACC through allocation TG-MR190035. This research was supported in part through computational resources and services provided by Advanced Research Computing Technology Services (ARC-TS), a division of Information and Technology Services (ITS) at the University of Michigan, Ann Arbor. The authors thank Prof. William Curtin from Brown University for insightful discussions.

\section*{Conflict of interest}
The authors declare that they have no conflict of interest.

\section*{Data and code availability}
The code with the implementation of Strain-boost Hyperdynamics within LAMMPS is available at \cite{LAMMPS_Hyperdynamics}. Data will be made available on reasonable request. 

\section*{Supplementary Materials}
Video S1: Dynamics evolution of the 3D dislocation slip trajectory corresponding to Fig.~\ref{fig:SlipPath_50nm_NbMo_1p0e4} \\ % SlipPath_Nb50Mo50_1p0e04_PositionY.mp4
Video S2: Dynamics evolution of the 3D dislocation slip trajectory corresponding to Fig.~\ref{fig:SlipPath_50nm_NbMo_1p0e3} % SlipPath_Nb50Mo50_1p0e03_PositionY.mp4

\newpage

%\section*{References}
\bibliographystyle{unsrt}  
\bibliography{references,references_extra}

\end{document}